\documentclass[useAMS,usenatbib,referee]{mn2e}

\usepackage{epsfig}
\usepackage{graphicx}
\usepackage{lscape}
\title[A non-stationary emission model  for LAT pulsars]
{Probing gamma-ray emissions of  $Fermi$-LAT pulsars with  
 a non-stationary outer gap model}
\author[J. Takata, C.W. Ng, and K.S. Cheng]{J. Takata $^{1}$\thanks{E-mail:takata@hust.edu.cn}, C.W. Ng$^{2}$\thanks{E-mail:ngchowing@gmail.com} 
and  K.S. Cheng$^{2}$\thanks{E-mail:hrspksc@hku.hk} \\
$^{1}$ School of physics, Huazhong University of Science and Technology, Wuhan 430074, China \\
$^{2}$ Department of physics, The University of Hong-Kong, Hong-Kong
}

\begin{document}

\date{}

\pagerange{\pageref{firstpage}--\pageref{lastpage}} \pubyear{2002}

\maketitle

\label{firstpage}

\begin{abstract}
We explore a non-stationary outer gap scenario for gamma-ray emission process
in pulsar magnetosphere. Electrons/positrons that migrate along the magnetic
field line and enter the outer gap  from the outer/inner boundaries
activate the pair-creation cascade and high-energy emission process.
In our model,  the rate of the particle injection at the gap boundaries   
 is key physical quantity to control  the gap structure and 
 properties of the gamma-ray spectrum.
 Our model assumes that  the injection rate  is  time variable  and 
 the observed gamma-ray spectrum are superposition of the emissions 
 from different gap structures with different injection rates
 at the gap boundaries. The calculated spectrum superposed
 by assuming power law distribution of 
the particle injection rate can reproduce sub-exponential cut-off feature in the 
gamma-ray spectrum observed by Fermi-LAT.  We fit  
the phase-averaged spectra for 
 43 young/middle-age pulsars and 14 millisecond pulsars with the model. 
 Our results imply that (1)  a larger particle injection at the gap
  boundaries  is more frequent 
 for the pulsar with a larger
 spin down power and (2)  outer gap with an injection rate 
 much smaller than the Goldreich-Julian value produces observed
 $>10$GeV emissions.  Fermi-LAT gamma-ray pulsars show
 that (i) the observed gamma-ray spectrum  below cut-off energy 
tends to be softer for the pulsar with a higher spin down rate 
and (ii) the second peak is more prominent in higher energy bands. 
 Based on the results of the fitting, 
we describe possible theoretical interpretations
 for these observational properties. 
 We also briefly discuss  Crab-like millisecond 
pulsars that show phase-aligned radio and gamma-ray pulses.
\end{abstract}


\section{Introduction}
\label{intro}
The Fermi gamma-ray telescope (hereafter, Fermi) launched in 2008 has 
facilitated the study of gamma-ray emission process in the pulsar
 magnetosphere. The Large Area Telescope on-board the Fermi (hereafter 
Fermi-LAT)  has measured the gamma-ray emissions from more 
than 150 pulsars (Fermi collaboration 2015), and  has measured
 the spectra and the pulse profiles above 1GeV with 
unprecedented sensitivity. For example, Fermi-LAT found that 
the gamma-ray flux above 
the cut-off energy at around $\sim 3$GeV decays  
slower than  pure exponential function (Abdo et al. 2010a, 2013). 
This cut-off behaviour  favours the emissions from the 
outer magnetosphere (e.g. slot gap, outer gap and annular gap) 
and rules out the classical polar cap scenario, 
which predicted a super exponential 
cutoff feature in the GeV spectrum because of the magnetic 
pair-creation process. Among Fermi-LAT pulsars, 
 20 pulsars are found  to show pulsed emissions  
in the energy range $>10$ GeV, including 12 up
 to $>25$ GeV (Ackermann et al. 2013) and their spectra clearly indicate   
sub-exponential cut-off features  above the cut-off energy 
 (Ackermann et al. 2013). The pulsed gamma-ray emissions from 
the Crab pulsar show single power law spectrum above  cut-off energy
 ($\sim 5$GeV) and extends to TeV energy bands 
 (Aleksi$\rm{\acute{c}}$ et al. 2011, 2012, 2014; 
Aliu et al. 2008, 2011, Abdo et al. 2010b). 
The GeV/TeV emissions from the Crab pulsar disagree with the spectra of 
the standard curvature radiation scenario 
(e.g. Cheng et al. 2000; Takata \& Chang 2007; Harding et al. 2008),
 and will originate from the inverse-Compton scattering process in the outer 
magnetosphere (Lyutikov et al. 2012; Harding and Kalapotharakos 2015) or 
pulsar wind region (Aharonian et al. 2012).  
We (Leung et al. 2014) reported the detection of 
 the pulsed emissions above 50GeV from the Vela pulsar, and showed that 
the previous models (e.g. Hirotani  2007; Takata et al. 2008) 
predicted a smaller flux level  at 50-100GeV energy
 bands than the observed flux.   
A study of sub-exponential spectrum above cut-off energy will 
discriminate among  emission models. 

In addition to sub-exponential cut-off behaviour, the
 Fermi-LAT observations have revealed several interesting relations between 
the gamma-ray emission properties and the spin down characteristics; 
 (1) the gamma-ray emission efficiency, 
which is the luminosity divided by the spin down power, 
 decreases with the spin down power, 
and (2) the spectrum between 100MeV 
and the cut-off energy at around $\sim 1$GeV 
tends to be softer for a larger spin down pulsars (Abdo et al. 2013), (3) 
 the second peak 
in the light curve is  in general more prominent in higher 
energy bands (e.g. Crab, Vela and Geminga pulsars, Abdo et al. 2013), and 
(4) Fermi-LAT millisecond pulsar with a higher spin
 down power and a larger magnetic field strength 
at the light cylinder tends to have Crab-like pulse profiles, in which 
radio/X-ray/gamma-ray pulses  are in phase (Ng et al. 2014).  
Explanations for these observed  properties with a model  will advance 
in understanding of the nature of the high-energy emission process in 
the pulsar magnetosphere.

The cause  of the formation of the non-exponential cut-off decay is still in 
debate. Abdo et al. (2010b) and Vigan$\rm {\grave{o}}$ and Torres (2015) 
argued that a sub-exponential cut-off 
in the observed spectrum could be understood as 
the superposition of several power law plus exponential cut-off functions with 
varying the photon index and the cut-off energy, for which  
the different components are produced at 
 the different region of the pulsar magnetosphere cutting 
 across our line of sight.  The contribution of the inverse-Compton scattering process (likewise the Crab
 pulsar) is one of the proposed models to explain the high-energy 
tail of the Vela pulsar (e.g. Lyutikov et al. 2012).  However, 
the  required soft-photon number density in the magnetosphere 
to explain the observed GeV flux level will be  much larger than one inferred
 from the optical/UV/IR observations of the Vela pulsar (Takata et al. 2008).

We (Leung et al. 2014) discussed the formation of the  spectrum of 
the Vela pulsar within framework of the outer gap model, and proposed 
a non-steady model. In this new outer gap model, 
the electrons and positrons that enter
the gap from outer and inner boundaries, respectively,  control 
the gap structure (size, particle distribution and electric field
strength etc.) and a smaller rate of the particle injection
produces thicker outer gap and
harder spectrum. The model suggested that the injection rate 
much smaller than Goldreich-Julian value  produces the observed 
gamma-ray emissions above 10GeV.  We argued that the rate of the
particle injection at the gap boundaries could
 fluctuate  with  time and  the observed gamma-ray
 spectrum is superposition of the emissions 
 from different  stationary gap structures with different injection rates.

In this paper,  we will discuss a detail of the three-dimensional 
calculation method for this new outer gap model,
 since we did not provide it in our previous observational 
paper (Leung et al. 2014). 
In section~\ref{vela}, we present the predicted spectrum and light curve
 of the Vela pulsar. We will discuss the observed energy dependent
  light curve.  In section~\ref{disc},  we will apply our model  
to  other gamma-ray emitting pulsars,
 and will discuss how our model interprets the observed 
 relation between  the spectral softness
 below cut-off energy  and the spin down power. 
We will also discuss the Crab-like millisecond pulsar
 and the limit of our model. 

\section{Theoretical model}
\label{magneto}
\subsection{Pulsar magnetosphere with outer gap accelerator}
 The global simulations have been developed to investigate structure
 of the magnetosphere with the high-energy emission region. Earlier particle 
simulations showed that the magnetosphere with no-pair-creation process
 settles down into a quiet state with electron cloud above the
polar caps, a positively charged  equatorial disc and vacuum gaps 
in the middle latitudes (Krause-Polstorff \& 
 Michel 1985; Smith et al. 2001; Wada \& Shibata 2007). Recent 
particle-in-cell simulations have shown the pulsar magnetosphere with the 
discharged particles created by the pair-creation process. Chen \& Beloborodov 
(2014) discussed that if pair-creation multiplicity is very high 
at outer magnetosphere around the light cylinder, the outer gap around 
the light cylinder was quenched and the magnetosphere is 
similar to the force-free solution with a super Goldrecih-Julian
 current sheet and the Y-point near 
the light cylinder, where are main high-energy emission 
region (Spitkovsky 2006). On the other hand, it is  also suggested that if
the pair-creation process in the outer magnetosphere is low,   
the outer gap can survive from the fill of discharge particles and it can be
 high-energy emission regions (Wada \& Shibata 2007; Yuki \& Shibata 2012). 
It is still under debate for the structure of  pulsar magnetosphere as well as 
the high-energy emission region, since the current global 
simulations are difficult to deal with the 
realistic pair-creation process by taking into account the position dependent 
 mean free path and soft-photon density. 

In this paper, we assume that the pulsar magnetosphere has an outer gap and 
the high-energy gamma-rays are produced by the curvature radiation process 
of the discharge pairs inside the outer gap. Our local model precisely
 calculate the pair-creation rate in the outer magnetosphere. As we will 
see in section~\ref{pair}, the optical depth of the photon-photon 
pair-creation process around light cylinder is of order 
of $\tau_{X\gamma}\sim 10^{-3}$ for most of pulsars, and 
 most of the gamma-rays emitted from  outer gap 
can escape from the light cylinder.

\subsection{Particle injection at the gap boundaries}
\label{injection}
To activate the gamma-ray emissions and subsequent pair-creation cascade in
the outer gap, the  charged particles (electrons and/or positrons) should
enter the gap along the   magnetic field line from outside the gap; the outer
gap will be inactive without the injection of the particles at the gap boundaries. In this paper,  we use
terminology ``injected current'' to refer
the electric current component carried by the
electrons/positrons that enter the outer
gap from the gap boundaries. The outer gap thickness in  the poloidal
plane affects
to the magnitude of the accelerating  electric field and
therefore hardness/luminosity of  the curvature emissions;
a thinner outer gap produces a smaller accelerating electric field 
and a softer/fainter  gamma-ray emissions. 
From electrodynamical point of view,  we expect that the outer gap has a  thickness
so that the pair-creation cascade in the gap  produces an electric
current of order of the Goldreich-Julian value
and hence the gap structure will be affected by amount of the particles (i.e. injected current) that
 enter the gap from the inner and/or outer boundaries.  Takata et al. (2006) calculated two-dimensional  
outer gap structure and investigated the dependency of 
 gamma-ray spectra on injection rates of the particles at the inner and outer boundaries.  
 They demonstrated that a larger injection 
 produces in general a thinner outer gap and a softer gamma-ray spectrum. For the
 inclination angle less than $\alpha<90$degree, the positrons and electrons can enter
 the gap from inner and outer boundaries, respectively.

 The physical origin of the injected particles  at
  the gap boundaries are argued as follows. 
As suggested by Shibata (1991, 1995), the  polar cap accelerator,
outer gap region, and  the pulsar wind region, where the electric
current crosses
the magnetic field lines, should be connected by the current circulating
the magnetosphere. As shown in global simulations (e.g. Yuki \& Shibata 2012),
we expect that the pair-creation
process in the polar cap accelerator will make the current that flows higher
latitude around the magnetic pole,
while the discharged particles  in the outer gap accelerator are
main current carriers at  lower-latitude region around the last-open field
lines.  The polar cap accelerator model usually assumes a particle injection
from the neutron star surface. For the inclination angle $\alpha<90^{\circ}$,
the electrons from the stellar surface are injected
into the polar cap accelerator and initiate the pair-creation cascade
process through the magnetic pair-creation and/or photon-photon pair-creation
processes (Daugherty \&  Harding 1996). The discharged pairs form the current
that flows higher altitude. Most of  particles from the polar cap region will
flow out from the magnetosphere and will form the pulsar wind. But it
has been suggested that some of negative particles (for $\alpha<90$degree)
cross  the magnetic field lines  towards equator 
due to ${\bf F}_{rad}\times {\bf B}$ drift
(Wada \& Shibata 2011l; Yuki \& Shibata 2012), where ${\bf F}_{rad}$ is the
radiation drag force, and they eventually return to the star along the
magnetic field lines at the lower latitude.  It is probable that on the way
from the light cylinder to the star, the returning electrons enter to
the outer  gap along the magnetic field line from the outer boundary.
The high-energy emissions by the returning electrons and subsequent
pair-creation cascade processes produce the discharged pairs  that
also contribute to the current flowing lower latitude around the last-open field lines.
In section~\ref{cont} we discuss the current conservation along the magnetic field line.

We can argue several possibilities for
the physical origin of the  positrons that enter the gap from  the inner boundary. 
In the polar cap accelerator, the discharged positrons will return
to the polar cap region.  If the star continuously absorbs the positrons
 more than electrons, it would be charged up positively.
To keep the charge of the star at constant, the positrons should
be re-emitted from stellar surface along the magnetic field lines
outside polar cap accelerator. Such positrons  could enter the outer gap
from the inner boundary and contribute to the electric current flowing along
the magnetic field lines that penetrate the outer gap. 

Moreover, the gamma-rays produced in  the outer gap will create
more pairs around the inner boundary  (c.f. Figure~\ref{map}),
and residual electric field could separate the charge particles.
These discharged pairs could effectively become the injection current at the inner boundary, because the main emission region of the outer gap is beyond the null charge surface. Takata et al. (2010)
also argued that the gamma-rays emitted towards the stellar surface by the incoming
particles may generate new pairs via the magnetic pair-creation process near the
stellar surface, and  some new pairs could be returned to the outer gap
due to complicated surface magnetic field structure. These
returning positrons also could enter  the gap from the inner boundary.

\subsection{Outer gap with time dependent particle injection}
Although the pulsed radio emissions averaged over
longer time-scale is stationary,
there is a wide range of variability in a shorter times scale in
the radio emissions from the pulsar (e.g. Kramer et al. 2002;
Lyne et al. 2010; Keane 2013). The micro-second variations seen in single pulse could be
produced by spatial fluctuation in the emission region. The pulse-to-pulse
variations on the timescale of millisecond to second likely represent
timescale of the temporal variation of the structure of
the emission region (e.g. time dependent pair-creation process/particle emissions
from the stellar surface).  The
longer timescale (second to year) variations associated with the mode
switching and nulling, which sometimes accompany
the variations of the spin down rate, could be related with the changes
of entire magnetosphere. These  observations suggest that the switching
between different states of magnetosphere is probably a general feature
of the pulsars
 
In this paper, we assume that the outer gap structure is temporal variable
and that the observed gamma-ray emissions are superposition of the different
outer gap structures. We argue that the non-stationary behaviour of the outer
gap is caused by the time variation of the  rate of particle injections  at the gap
boundaries. We expect that the time-scale of  variations is  of order of or longer than
the  crossing timescale of the light cylinder, $\tau_c\sim R_{lc}/c=P_s/2\pi$, where 
$R_{lc}=cP_s/2\pi$ is the light cylinder radius and $P_s$  is the pulsar 
spin period.  For example, as we discussed above,
the discharged pairs produced around the inner boundary could effectively
become the origin of injected particles  at inner boundary.  In such a case,
the temporal variation  of the outer gap will be related to 
the variation of the pair-creation rate around the null charge surface.
Since the pair-creation
rate depends on the gamma-ray intensity   and surface X-rays intensity,
which is affected  by the returning particles  (c.f. section~\ref{xray}),
around the light cylinder, the expected time-scale of the variation will
be  of order of the light-cylinder crossing time-scale $\tau_c$. 
The variation of the  electrons  returning
from the pulsar wind region, which  will enter the gap from the outer boundary,
will be of order of or longer than the light-cylinder
crossing time-scale, since the time-scale shorter than the crossing time-scale
may be smoothed out during the travel around global magnetosphere.

We assume that the observed gamma-ray
spectrum is a superposition of the emissions from various
{\it stationary} gap structures with various particle injection rates at the gap
boundaries, and  the stationary outer gap structure for an injection rate  forms
with the crossing time-scale $\tau_c$.
For a fixed particle injection rate, our stationary solution will be stable for 
a small perturbation. For example, if the accelerating electric field increase 
from the stationary solution, the curvature photon energy and hence 
pair-creation rate increase from the stationary solution. The increase 
of the number of pairs try to screen the perturbed electric field.  

\section{Basic Equations}
In this section, we describe our basic equations for solving the gap structure
with  a fixed  injection rate at the boundary. By using vacuum rotating
dipole magnetic field,
we solve the Poisson equation to obtain the accelerating electric
field (section~\ref{accele}), for which the charge density in the gap is obtained
by solving the continuity equations for the electrons and positrons (section~\ref{cont})
with the curvature radiation process and pair-creation process (section~\ref{casc}).
In section~\ref{cont}, we will discuss the conservation of the electric
current along the magnetic
field line.
\subsection{The accelerating electric field}
\label{accele}
We investigate the outer  gap structure  under the steady condition 
that $\partial_t+\Omega\partial_\phi=0$ with $\Omega$ being spin angular
 frequency.  The electric field along the magnetic field line arises in 
the charge depletion region from so called Goldreich-Julian charge
density, and it  accelerates 
 the positrons and electrons to an ultra-relativistic 
speed. The Poisson equation for the accelerating electric field is written as 
\begin{equation}
\triangle\Phi_{nco}=-4\pi(\rho-\rho_{GJ}),
\label{poisson}
\end{equation}
where $\rho$ is the space charge density
 and $\triangle$ is the  Laplacian. In addition, 
$\rho_{GJ}=-\Omega B_z/2\pi c$ is the Goldreich-Julian charge density,
 where $B_z$ is the component of the magnetic field projected to the 
rotation axis. The accelerating electric field along the magnetic 
field line is computed from $E_{||}=-\partial \Phi_{nco}/\partial s$, 
where $s$ is the distance along the magnetic field line.  

To solve the Poisson equation (\ref{poisson}), we adopt 
coordinate system based on the distance along the field line, $s$, from
the star ($s=0$) and the magnetic coordinates,
 $\theta_*$ and $\phi_*$, which are angles measured 
from and around  the magnetic axis, respectively  (Hirotani 2006). 
We define $\theta_*=0$ at  the north  magnetic pole   
and $\phi_*=0$ (magnetic meridian) at the plane that includes  
 the rotation axis and north magnetic pole for inclined rotator. 
 The  coordinates $(s~,\theta_*,~\phi_*)$ relate with the canonical spherical
 coordinates $(r,~\theta,~\phi)$, for which 
$z$ axis coincides with the rotation axis, as 
\begin{equation}
r=R_s+\int_0^s\frac{B_r}{B}ds, 
\end{equation}
\begin{equation}
\theta=\theta_0(\theta_*,\phi_*)+\int_0^s\frac{B_{\theta}}{rB}ds, 
\end{equation}
and  
\begin{equation}
\phi=\phi_0(\theta_*,\phi_*)+\int_0^s\frac{B_{\phi}}{r\sin\theta B}ds, 
\end{equation}
 where $R_s$ is the stellar radius, $B$ is the local magnetic field strength 
 and $(B_r,~B_{\theta},~B_{\phi})$ are $(r,~\theta,~\phi)$ components of 
the magnetic field, respectively. We define that $\theta_0=0$ corresponds to 
the rotation axis and $\phi_0=0$ is the magnetic meridian. 
 We can relate between 
($\theta_0,~\phi_0$) and  ($\theta_*,~\phi_*$) as 
  $\cos\theta_0(\theta_*, \phi_*)=\cos\theta_*\cos\alpha-\sin\theta_*
\cos\phi_*\sin\alpha$ and  $\cos\phi_0(\theta_*,\phi_*)
=(\sin\theta_*\cos\phi_*\cos\alpha+\cos\theta_*\sin\alpha)/\sin\theta_0$
with $\alpha$ being the inclination angle.

\subsubsection{Boundary conditions}
For 3-D outer gap, there are  six boundaries, that is, inner (stellar side), outer (light cylinder side),
lower, upper, leading side and trailing side boundaries. For inclined rotator, the charge deficit
 region at  the azimuthal angle  $|\phi_*| >100^{\circ}$ 
is in general   less active,  because the null charge surface is located 
 close to the light cylinder, and because the electric
 field is too small to  boost the charge particles up to ultra-relativistic 
speed that can produce  the high-energy  gamma-rays. 
In this paper, therefore,  we put the numerical boundaries  on 
the magnetic field lines labelled by 
  $\phi_*=\pm 100^{\circ}$  for  the leading side
(positive sign) and the trailing side (negative sign)  
of the gap, and impose  the mathematical boundary conditions 
that $\Phi_{nco}=0$. 

For fixed azimuthal angle $\phi_*$,
the lower and upper gap boundaries lay on the magnetic field lines.
We fix the lower boundary at  the last-open field lines
and impose $\Phi_{nco}=0$ on it. We also impose $\Phi_{nco}=0$ on the upper boundary and
solve it's position, for which the gap can create  an assumed  electric
current density
(c.f. section~\ref{model}). In the calculation, we  set the outer boundary near 
the light cylinder and impose $E_{||}=0$ on the boundary.  We initially apply  
the numerical boundary at  $s\sim 1.5R_{lc}$ and solve the gap dynamics. 
If the electric field changes its sign around the given 
outer  boundary, then  we set 
new outer boundary at the location where the solved electric field changes its
sign, because  we anticipate that the outer gap should be
 unstable if the field-aligned electric field changes its sign inside 
 the gap.

 Finally,  let us consider the inner
 boundary (stellar side). Because we assume that
 there is no potential drop  between the stellar
surface and the inner boundary, we impose the conditions $\Phi_{nco}=0$ and
$E_{||}=0$.  Since arbitrary given boundary does not satisfy 
 both  the Dirichlet-type and the  Neumann-type conditions,  we seek for
 the appropriate boundary  by moving the boundary step by step. 
With two-dimensional analysis,
Takata et al. (2004) discussed that the inner boundary of the outer gap 
starts from the position where the charge density of the current
carriers is equal to the Goldreich-Julian charge density. For example,
the outer gap starts from the null-charge density of the
Goldreich-Julian charge density, if the gap is vacuum.  
On the other hand, the inner boundary will locate on the stellar surface,  
if the electric current created inside the gap
is $j_{gap}\sim \cos\alpha$  in units of
the Goldreich-Julian value, $\Omega B/2\pi$.

\subsection{Continuity equations }
\label{cont}

In this paper, we assume that the inclination angle of the magnetic pole is
less than 90 degree. In such a case, the positive electric field along
the magnetic field line accelerates
the positrons towards light cylinder and electrons towards the stellar surface,
respectively. In the outer gap, we can anticipate that new born  pairs in
the gap are immediately   charge separated and are boosted
to ultra-relativistic  speed by the electric field along the magnetic field
line (Hirotani \& Shibata 1999); that is,
we can assume that all  positrons and electrons in the outer gap
move towards the light cylinder and towards the stellar surface, respectively, 
  with the speed of light. Under these conditions,
  the continuities of the number density
of the positrons (plus sign) and
of the electrons (minus sign)  may be written as 
\begin{equation}
\frac{d}{ds}\left(\frac{cN_{\pm}}{B}\right)=
\pm S(s,\theta_*,\phi_*), 
\label{conteq}
\end{equation}
where $S(s,\theta_*,\phi_*)$ is the source term due to photon-photon 
pair-creation process.  The electric current density per magnetic
flux tube in the gap
is given by $ce(N_+(s)+N_-(s))/B$.  Fixing ($\theta_*,~\phi_*$),
  the continuity equation~(\ref{conteq})
   satisfies the current conservation along the field line, that is, 
\begin{equation}
  j_{tot}\equiv
  ce \frac{N_+(s)+N_-(s)}{\Omega B/2\pi}={\rm constant~along}~s 
\end{equation}
where we normalized the current density by the local Goldreich-Julian value.
We  define the normalized current densities carried by the positrons and
electrons as
\[
j_{\pm}(s)\equiv ce \frac{N_{\pm}}{\Omega B/2\pi},
\]
and $j_{tot}=j_{+}(s)+j_{-}(s)$.
With the equation~(\ref{conteq}),  the solutions for $j_{\pm}$ can be written
as 
\[
j_+(s)=j_{in}+\int_{s_{in}}^{s}S'(s')ds',
\]
and 
\[
j_-(s)=j_{out}+\int_{s}^{s_{out}}S'(s')ds',
\]
respectively, where $S'(s)=2\pi eS(s)/\Omega$, $s_{in}$ and $s_{out}$ represent
the positions of the inner boundary and outer boundary, respectively, and the injection current
$j_{in}$ (or $j_{out}$) represents  number
of positron (or electron) that enters the gap from the inner (or outer)
boundary per unit time per unit area  and per  magnetic flux tube.
The origin of the particles injected into the gap
were discussed in section~\ref{injection}.  In terms of
($j_{in},~j_{out},~j_{gap}$),
the conservation of the electric current along the magnetic field line
becomes
\begin{equation}
  j_{tot}=j_+(s)+j_-(s)=j_{in}+j_{out}+j_{gap},
  \label{cons}
\end{equation}
where
\[
j_{gap}=\int_{s_{in}}^{s_{out}}S'(s')ds'
\]
which represents the current component carried by the created pairs in the outer gap
(hereafter we use terminology ``gap current'' to refer $j_{gap}$).
Equation~(\ref{cons}) tells us  that the electric current
along the magnetic field line
is  sum of the injection currents at gap boundaries  plus gap current.
We note that as long as  the current flows along the magnetic field line
that penetrates the outer gap, the magnitude of current density per magnetic flux tube
is equal to $j_{tot}=j_{in}+j_{out}+j_{gap}$ both outside and inside the gap.  Hence
there is no current discontinuity along the magnetic field line.
To close the current circuit, the trans-field current flow should
appear in somewhere beyond the light cylinder (Shibata 1991, 1995). In this
paper, since the structure of the magnetosphere outside the light cylinder is
beyond out of scope,  we just assume that the cross-field region is far from
the outer boundary, and that the injected  electrons cross the outer boundary
along the magnetic field line.

Actual values
for the total current $j_{tot}$,
injected currents $j_{in}$ and $j_{out}$
should be solved with the complicated physics (e.g. energy-angular loss relation
among the polar cap accelerator, outer gap and pulsar wind region, Shibata 1991)
of the global pulsar magnetosphere.  For example, the injection current $j_{in}$ might
be solved together with the outer gap activity and positron re-emission from the
neutron star surface, which is related to the charge redistribution over the polar
cap region. As we discussed in section~\ref{injection},
the injection current $j_{out}$ at the outer boundary will be related to the
physics of the formation of the pulsar wind. The total
current $j_{tot}$ running through the outer gap should be solved with
global pulsar magnetosphere including the polar cap, outer gap and pulsar wind region.
Because of the large theoretical
uncertainties of the global structure, however,  our local model
treats ($j_{tot},~j_{in},~j_{out}$) or $(j_{gap},~j_{in},~j_{out})$  as a set of
the free parameters. In section~\ref{model}, we describe how our model assumes
the values of  $(j_{gap},~j_{in},~j_{out})$.

\subsection{Curvature radiation and pair-creation processes}
\label{casc}

To calculate the source term in equation~(\ref{conteq}), we 
compute the pair-creation process between the gamma-rays emitted  by 
the curvature radiation  and thermal radiation from the stellar surface. We
calculate the Lorentz factor of the accelerating electric field by
assuming force balance between the acceleration force  and the 
back reaction  force of the curvature radiation process as 
\begin{equation}
\Gamma=\left(\frac{3R_c^2E_{||}}{2e}\right)^{1/4},
\label{gamma}
\end{equation}
where $R_c$ is the curvature radius of the magnetic field line.  
The number of curvature photons emitted 
per unit time from the particle with a Lorentz factor $\Gamma$ is 
\begin{equation}
  P_c=\frac{8\pi}{9}\frac{e^2\Gamma}{hR_c}=3.2\times10^6
  \left(\frac{\Omega}{100{\rm s^{-1}}}\right)\left(\frac{\Gamma}{10^7}
\right)\left(\frac{R_{c}}{R_{lc}}\right)^{-1}\ \ \mathrm{s^{-1}}. 
\label{photons}
\end{equation}
The spectrum of the curvature radiation from the  particle is described 
by  
\begin{equation}
\frac{dN_{\gamma}}{dE_{\gamma}}=\frac{\sqrt{3}e^2\Gamma}{hR_cE_{\gamma}}
F(x), 
\end{equation}
where $x\equiv E_{\gamma}/E_c$ with  
\begin{equation}
E_c=\frac{3}{4\pi}\frac{hc\Gamma^3}{R_c}=0.1\left(\frac{\Omega}
{\mathrm{100s^{-1}}}\right)\left(\frac{\Gamma}{10^7}
\right)^3\left(\frac{R_{c}}{R_{lc}}\right)^{-1} \ \ \mathrm{ GeV}, 
\end{equation}
and 
\begin{equation}
F(x)=\int_x^{\infty}K_{5/3}(y)dy, 
\end{equation}
where $K_{5/3}$ is the modified Bessel function of the order of $5/3$.

The emitted curvature photons may convert into 
  new electron and positron pairs  through  the pair-creation process. 
The mean free path of the pair-creation $l_p$ is 
\begin{equation}
\frac{1}{l_p}
=(1-\cos\theta_{X\gamma})c\int_{E_{th}}^{\infty}dE_{X}\frac{dN_X}
{dE_X}(\mathbf{r},E_{X})\sigma_p(E_{\gamma},E_{X}),
\label{meanf}
\end{equation}
 with  $dE_{X}\cdot dN_X/dE_{X}$ being  the X-ray number
density between energies $E_{X}$ and
$E_X+dE_X$, $\theta_{X\gamma}$ the collision angle
between an X-ray photon and a gamma-ray photon, and 
$E_{th}=2(m_ec^2)^2/(1-\cos\theta_{X\gamma})E_{\gamma}$  the
threshold X-ray energy for the  pair creation. In addition, 
 the pair creation cross-section $\sigma_p$  is given by
\begin{equation}
\sigma_{p}(E_{\gamma},E_{X})=\frac{3}{16}
\sigma_{T}(1-v^2)\left[(3-v^4)\ln\frac{1+v}{1-v}-2v(2-v^2)\right],
\label{cross}
\end{equation}
where
\[
v(E_{\gamma},E_{X})=\sqrt{1-\frac{2}{1-\cos\theta_{X\gamma}}\frac{(m_ec^2)^2}
{E_{\gamma}E_{X}}},
\]
and  $\sigma_{T}$ is the Thomson cross-section.  In this paper, we
consider the thermal  X-ray photons from the stellar surface. 
 At the radial distance $r$ from the centre of the star,
 the  photon number density between energy
$E_{X}$ and $E_{X}+dE_{X}$ is given by
\begin{equation}
\frac{dN_X}{dE_X}=2\pi\left(\frac{1}{ch}\right)^3
\left(\frac{R_{eff}}{r}\right)^2
\frac{E_X^2}{\exp(E_X/kT_s)-1},
\label{soft}
\end{equation}
where $R_{eff}$ is the effective radius and 
 $T_s$ refers to the surface temperature. 

\subsection{X-ray emissions from NS surface}
\label{xray}
\begin{figure}
\includegraphics{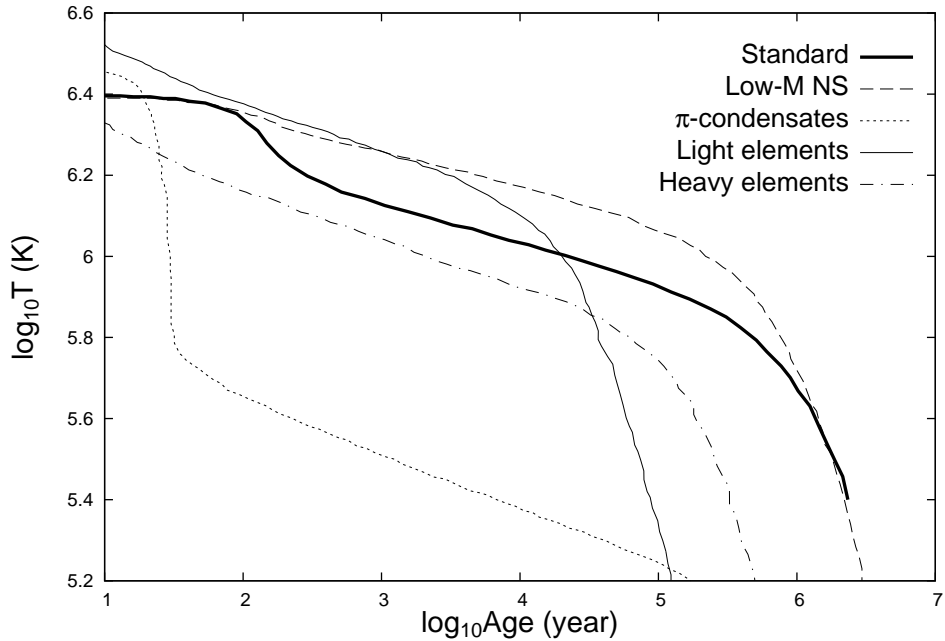}
\caption{Cooling curves for different neutron star models; standard  
model (thick solid line), low-mass neutron star model (dashed line), 
high-mass neutron star with $\pi$-condensate at the core (dotted line), 
the neutron star with light elements in envelop (thin-solid line), and 
the neutron star with heavy elements in envelop (dash-dotted line). The 
cooling light curves were taken from ${\rm\ddot{O}}$zel (2013).}
\label{cool}
\end{figure}

 In this paper, we consider two types of the surface emissions, namely, 
the neutron star cooling emissions and  heated polar cap emissions.
Both surface emission processes could contribute to the observed thermal 
X-ray emissions from the young/ middle-age pulsars
 (e.g. Caraveo et al. 2004 for the Geminga pulsar), while only   
heated polar cap emission should be important  for the millisecond pulsars
 (Zavlin 2007; Takata et al. 2012) 

 For the neutron star cooling model, the temperature as a function 
of the age  really depends on the neutron star model (Yakovlev \& Pethick 2004),
  as shown in Figure~\ref{cool}. Although the observations of 
 surface temperature could exclude the  high mass neutron star with 
$\pi$-condensate core model, it is still under debate for the 
 neutron star model. In this  paper, therefore, 
we use the cooling curve predicted by 
the standard  model (thick line in Figure~\ref{cool}),
 which will provide typical surface temperature 
for fixed age of the neutron star. We assume  the spin down age 
$\tau_s=P_s/2\dot{P}_s$, where $\dot{P}_s$ is the time derivative of
 the spin period, for the true age of  the young/middle-age pulsars 

For the  heated polar cap emissions, we apply the model 
 by Takata et al. (2012), in which the X-ray emissions from the heated polar 
cap region are composed of  two components, namely, a  core component and a 
rim component.  The core component shows a higher temperature but a 
smaller effective radius ($R_c\sim 10^{3-4}$cm), and the rim component has 
a lower temperature but larger effective radius ($R_r>10^{5}$cm). 
In their model, the bombardment of the returning pairs on the polar cap region 
causes the core component, while the irradiation of  
$\sim 100$MeV gamma-rays that are emitted near the stellar 
surface  by the returning particles heats up  the surface 
and produce the rim components. We expect that 
with a smaller effective area ($R_c\sim 10^{3-4}$cm), 
the core component does not illuminate the outer gap, while 
the rime component with the effective radius $R_r>10^5$cm has a greater 
likelihood of illuminating the outer gap. This model predicts the 
temperature of the rim component as (c.f. equation 21 in Takata et al. 2012)
\begin{equation}
T_r\sim 10^6{\rm K}\left(\frac{P}{1{\rm ms}}\right)^{-3/28}
\left(\frac{B_s}{10^{8}{\rm G}}\right)^{3/28}
\left(\frac{R_r}{10^{5}{\rm cm}}\right)^{-3/7}
\end{equation}
In this paper, we assume  $R_r\sim 4\times 10^5$cm to match with  
typical observed temperature and effective radius of the millisecond pulsars.

\subsection{Model parameters}
\label{model}
\begin{figure}
\includegraphics{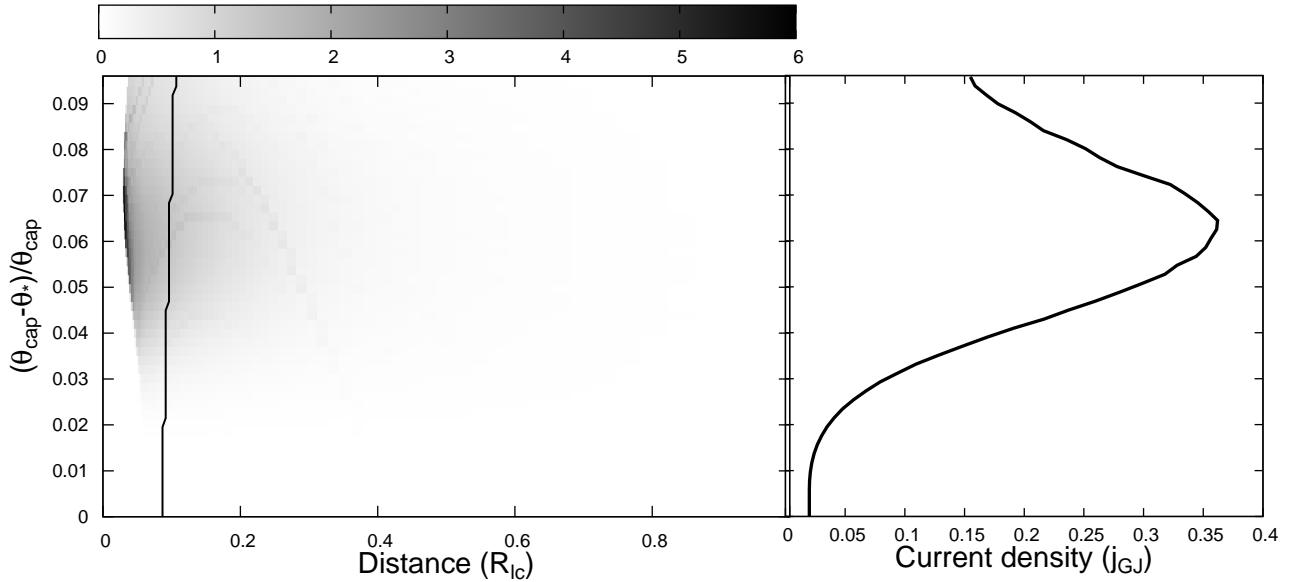}
\caption{Left:  The created electric current density per unit
  length, $dj_{gap}/ds$,
 in the gap at the magnetic meridian $\phi_*=0$. The abscissa axis is the 
distance in units of the light cylinder radius 
from the stellar surface along the magnetic field lines, and the vertical axis 
represents the gap height from the lower boundary ($\theta_*=\theta_{cap}$) 
 to the upper boundary.  The vertical thick line shows the position of the null charge points. Right: The trans-field distribution of the total current 
density in units of Goldreich-Julian value. The 
results are for the Vela pulsar with the injection 
rate $j_{ex}=10^{-2}$ and the inclination angle $\alpha=60^{\circ}$. }
\label{map}
\end{figure}

In the present model, we treat the injection currents 
$(j_{in},~j_{out})$ and  the gap current $j_{gap}$  as
the model parameters and they are relate to the total current
as  $j_{tot}=j_{in}+j_{out}+j_{gap}$ (c.f. section~\ref{cont}).
In addition,  the inclination angle $\alpha$ is model parameter and
we will fix the inclination angle $\alpha=60^{\circ}$ in this paper. Since
we focus on the observed phase-averaged  spectrum,  we do not introduce
the Earth viewing angle. By integrating the emissions from whole outer gap region, we will 
compare the model spectrum with the observed phase-averaged spectrum.

For the injection  currents $(j_{in},~j_{out})$ at the gap boundaries,
we  assume constant over the boundaries and we impose
also $j_{in}=j_{out}$ for reducing the model  parameters, that is, we assume
same particle injection rates at the inner and outer boundaries.
Choice of the equal injection rates at the gap boundaries 
is arbitrary, and it is not necessary for the real case. For the photon-photon
pair-creation process with the X-rays from the neutron star surface,
the gap structure is more sensitively to the choice of the injection 
current at the outer boundary. This is because the pair-creation 
process  between the gamma-rays emitted by the inward migrating electrons 
and the surface X-rays are head-on collision process, while the pair-creation 
process of the outward propagating gamma-rays from the positrons is tail-on 
collision process. Hence, most of the pairs are created by the inward
 propagating gamma-rays. We expect that if we assume no injection current 
at the outer boundary (i,e, $j_{out}=0$), the outer gap size will 
become thicker than the solutions discussed in this paper. We will study 
the gamma-ray emissions from the outer gap with $j_{in}\neq j_{out}$ in the 
subsequent papers.  Here we define  total injection current
$j_{ex}$, as
\[
j_{ex}\equiv j_{in}+j_{out},
\]
which is time variable quantity in our model.  In the model, we will
apply $10^{-8}<j_{ex}<0.1$ (see section~\ref{process}).

The gap current, $j_{gap}$  is limited  as follows.
Figure~\ref{map} represents an example of the calculated  gap structure in
the plane defined by $\phi_*=0$;
the left panel shows the photon-photon pair-creation rate in the 
gap and the right-hand panel shows the trans-field distribution of the total current
density ($j_{tot}$).
We can see in the figure that the calculated
gap current ($j_{gap}=j_{tot}-j_{ex}$)
increases as increase of the height from the lower boundary
(last-open field line), and that there is a maximum value of the gap current ($\equiv j_{gap,max}$).
  This is because the gamma-rays propagate in the convex side of the magnetic field lines. Around
the upper boundary, the gap current decreases because the electric field decreases
there and because the gamma-rays emitted at lower region  do
not illuminate upper region of the gap.
In the Figure~\ref{map}, the gap current therefore  becomes maximum
at around 70-80\% of the gap thickness,  and  the maximum
gap current is $j_{gap,max}(\phi_*=0)\sim 0.35$. 

In the paper, we teat $j_{gap,max}(\phi_*)$ as the model parameter.
We can find in Figure~\ref{map} that the position of the
inner boundary on the magnetic field line
that has a large gap current $j_{gap}$ shifts towards stellar surface from 
the GJ null charge surface.
As discussed in Takata et al. (2006), the inner boundary of the outer gap will
locate near the stellar surface, if the created current
is of order of $j_{gap}\sim \cos\alpha$. Within 
the framework of the calculation method, however, it is difficult to 
obtain such a stable solution, in which the field
aligned electric field does not change its direction, 
 if the gap current approaches to $j_{gap}\sim \cos\alpha$. In 
the calculation, hence, we 
assume the maximum gap  current $j_{gap,max}$ with a value slightly
 smaller than $\cos\alpha$. In the  model calculation,
 we  assume the inclination angle $\alpha=60^{\circ}$, and we solve the location 
 of the upper boundary so as to create the gap current
 of $j_{gap,max}(\phi_*)\sim 0.3-0.4$, which does not depend on the azimuthal angle (but see section~\ref{process}
  for large $|\phi_*|$).
 As long as  $j_{gap, max}\sim \cos\alpha$, the exact value of $j_{gap,max}$ does not 
affect much on the calculated gamma-ray spectra. 

\subsection{Calculation Process}
\label{process}

With the specified injection current, $10^{-8}<j_{ex}<0.1$,  and
fixed maximum gap current $j_{gap,max}=0.3-0.4$, we self-consistently solve the outer gap structure,
as follow.
We start the calculation by solving the Poisson equation (\ref{poisson})
for a {\it vacuum} outer gap
with a very {\it thin} thickness. Using the calculated electric field along
the magnetic field line, we calculate the terminal Lorentz factor (\ref{gamma})
at the each calculation grid. Given the injection current, $j_{ex}$,
we solve the continuity equation (\ref{conteq}) with the curvature radiation
and pair-creation processes,
and then we obtain new distribution of the charge density inside the gap. 
With the new charge distribution, we solve the Poisson equation to update the 
electric field, which subsequently modifies the charge density distribution. 
We iterate this procedure until all physical quantities converge.

If the gap thickness is too thin, the magnitude of
the electric field is not enough high to boost the
electrons/positrons up to  ultra-relativistic speed, and the pair-creation cascade
does not initiate in the gap.  As a next step, therefore, we increase the thickness
of the gap. For a  fixed magnetic azimuth, the gap current $j_{gap}$  has a
distribution in the direction of the latitude $\theta_*$, as  Figure~\ref{map}
indicates.   If the maximum current density at fixed magnetic azimuth 
is smaller than $j_{gap, max}\sim 0.3$ (for $\alpha=60^{\circ}$), 
we slightly increase the gap height. The increase of the gap height produces the 
increase of the accelerating electric field and results in the increase of
the gap current. We solve the outer gap dynamics with new 
upper boundary and  obtain new distribution of the gap current. Updating 
the gap upper boundary step by step, we finally obtain the desired gap 
structure for the fixed  injection  current $j_{ex}$. 

\begin{figure}
\includegraphics{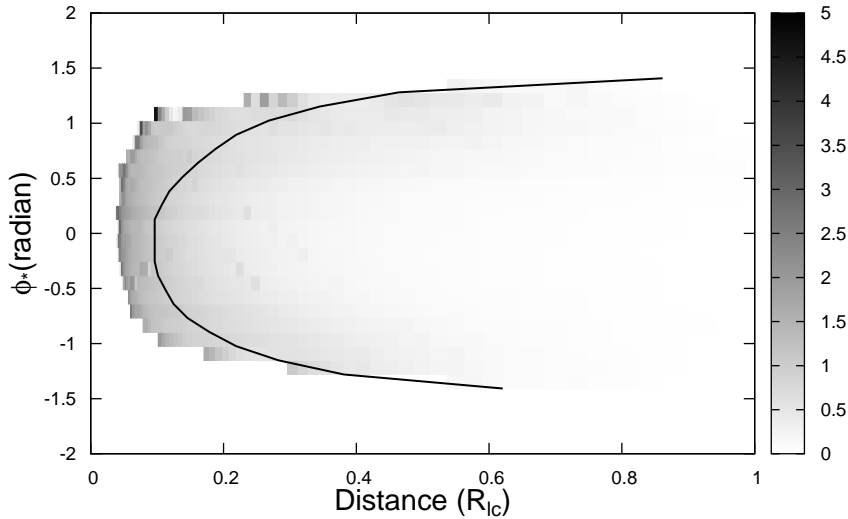}
\caption{The created current density, $dj_{gap}/ds$,
 at middle of outer  gap. The vertical axis 
represents the magnetic azimuth and $\phi_*=0$ represents magnetic meridian. 
The results are for the Vela pulsar, injection 
rate $j_{ex}=10^{-2}$ and the inclination angle $\alpha=60^{\circ}$.
 The thick line shows the position of the null charge points.}
\label{map1}
\end{figure}

In the model, the trans-field thickness of the outer gap
 is a function of  the magnetic azimuth $\phi_*$,  
and the gap thickness is the minimum   at around the magnetic meridian 
$\phi_*=0$. This is because the  gap thickness relates to the radial distance 
to the null charge point on the last-open field line from the stellar surface.
 At around $\phi_*=0$,  the radial distance to the null charge point
 becomes minimum (c.f. Figure~\ref{map1}), 
and hence the number density of the surface X-rays 
around the inner boundary of the gap becomes maximum. Because 
the pair-creation rate increases as increasing of the number density of 
the surface X-rays, the pair-creation rate inside the outer gap becomes 
maximum around the magnetic meridian. As a result, the gap thickness becomes 
minimum around the magnetic meridian. 

If the pair-creation rate is very low, the outer gap can become very thick. 
For example, on the magnetic 
field lines labeled by  the azimuthal angle  $|\phi_*|>90^{\circ}$, 
since the null charge point at the last-open field lines  are located near 
the light cylinder (c.f. Figure~\ref{map1}), the pair-creation 
rate is very low  and therefore the outer gap can become very thick. 
In  the calculation,  we set  possible maximum  thickness at  
 $\delta\theta_{*,max}(\phi_*)=0.8\theta_{cap}$, namely 80\% of the open field 
line region for the fixed $\phi_*$. For such an azimuthal angle,
the maximum gap current $j_{gap,max}(\phi_*)$ is smaller
than $j_{gap,max}<0.3-0.4$.  We note that there is  
critical magnetic field line for fixed $\phi_*$,  above which the null 
charge points on the magnetic field lines  locate 
outside the light cylinder, and therefore  a part of the outer gap 
 in the calculation would  locate outside right cylinder. 
 In the present calculation, we ignore the radiation process and the 
pair-creation process outside the light cylinder, (1) because the emissivity
 of the curvature radiation and the pair-creation rate 
will be very low compared to those inside the light cylinder and 
(2) because the special relativistic effect (e.g. retarded 
 electric potential)  and magnetic field bending due to the 
magnetospheric electric  current should be taken into account  to obtain 
the correct gap structure outside the light cylinder.   For the very 
thick outer gap, the pair-creation rate around 
the upper boundary is negligibly low  and most emissions are produced 
in  lower gap region.

Our model assumes that the observed gamma-ray spectrum is a superposition 
of the emissions from various gap structures 
with the various particle injection rates at the gap boundaries. 
Since our local model cannot determine the distribution 
of the injection rate, which will relate to  the physics in the 
 source region (e.g. polar cap),  we assume a  power-law distribution with  
\begin{equation}
\frac{dN_{ex}(j_{ex})}{d{\rm log_{10}}j_{ex}}=Kj_{ex}^{p},~~j_{ex,min}<j_{ex}<j_{ex,max},
\label{cdis}
\end{equation}
where $K$ is the normalization factor and it is calculated from 
$\int_{j_{ex,min}}^{j_{ex,max}}(dN_{ex}/d{\rm log_{10}}j_{ex})d{\rm log_{10}}j_{ex}=1$. 
We fix  the minimum injection rate at $j_{ex,min}=10^{-8}$, because  
the solved outer gap for the most pulsars has the maximum thickness,
 $\delta\theta_{gap}/\theta_{cp}=0.8$, for $j_{ex,min}=10^{-8}$.  We set 
the maximum injection rate at $j_{ex,max}=0.1$ so that the injection rate
 is smaller than created current in the gap $j_{gap}\sim 0.3_{GJ}$.  
With the function form  of equation~(\ref{cdis}), a larger (or smaller) 
injection current dominates in the distribution   
for the power-law index $p>0$ (or $p<0$). The superposed spectrum becomes 
\begin{equation}
F_{tot}(E_{\gamma})=\int_{j_{ex,min}}^{j_{ex,max}}F_{\gamma}(E_{\gamma})
\frac{dN_{in}}{d{\rm log_{10}}j_{ex}}d{\rm log_{10}}j_{ex},
\end{equation}
where $F_{\gamma}(E_{\gamma})$ is the gamma-ray spectrum for a fixed 
injected rate.

\section{Application to the Vela pulsar}
\label{vela}

\begin{figure}
\includegraphics{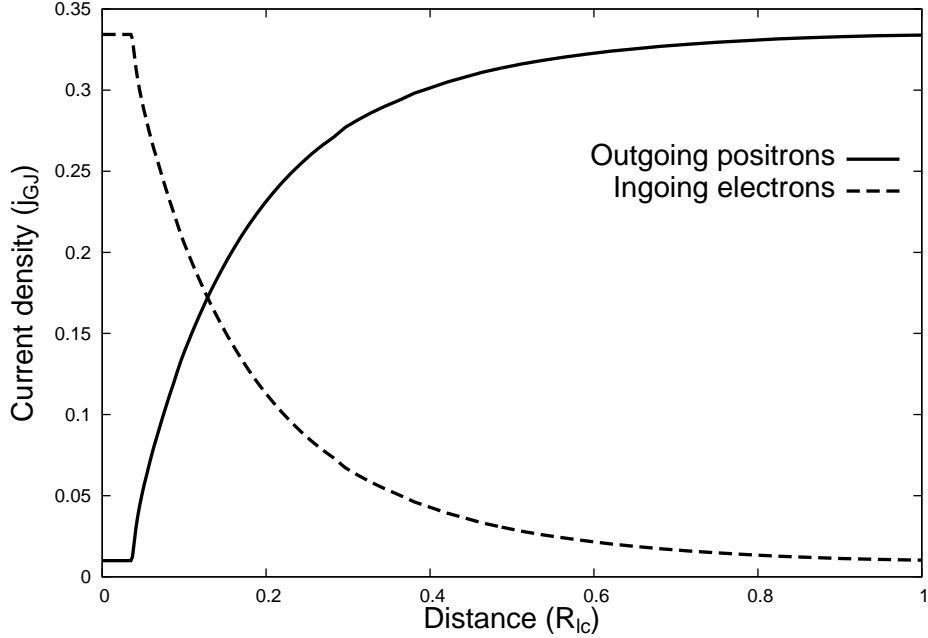}
\caption{Evolution of the electric current density along the magnetic field line, which penetrates  the outer gap accelerator.  The solid and dashed lines are for the outgoing positrons and for the in-going electrons, respectively. The results are for the Vela pulsar with injection rate $j_{ex}=10^{-2}$ and the inclination angle $\alpha=60^{\circ}$. }
\label{current}
\end{figure}

\begin{figure}
\includegraphics{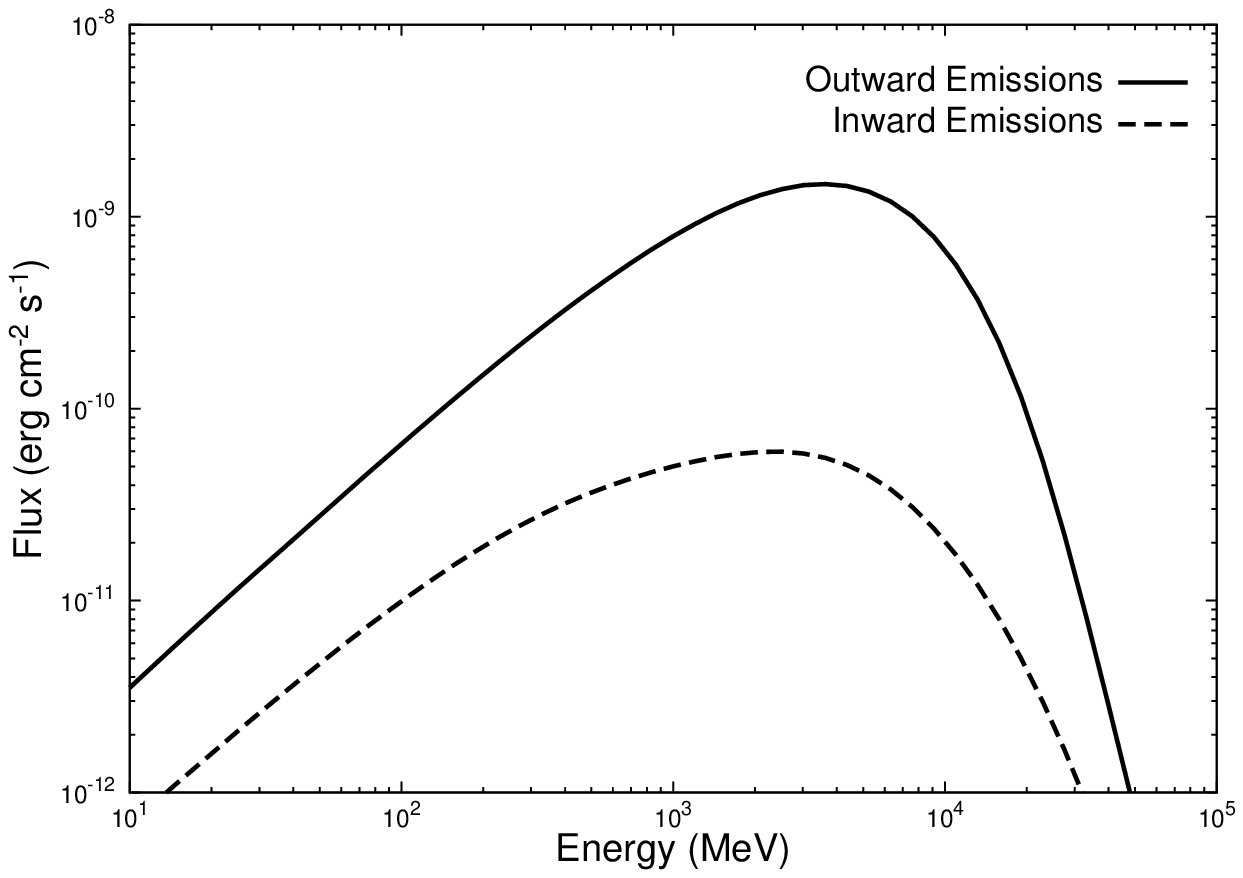}
\caption{Example of the gamma-ray spectra of the outwardly migrating particles
 (solid line) and of the inwardly migrating particles (dashed line). 
The results are for the Vela pulsar, injection 
rate $j_{ex}=10^{-4}$ and the inclination angle $\alpha=60^{\circ}$. }
\label{spec}
\end{figure}

In this section, we will apply the model to the Vela pulsar (PSR J0835-4510)
 and will discuss the general properties of the outer gap structure 
and the gamma-ray spectrum predicted 
by the model.

\subsection{Pair-creation in the gap}
\label{pair}
 Figures~\ref{map} and~\ref{map1} show the created gap current  (in units
 of the Goldreich-Julian current density) per unit length,
 $dj_{gap}/ds$, in the gap  and $\phi_*=0^{\circ}$ (magnetic meridian).
 In Figure~\ref{map}, the bottom ($\theta_*=\theta_{cap}$) 
represents the lower boundary and top ($\theta_*\sim 0.91\theta_{cap}$) 
is the upper  boundary of the gap. The results are for 
the injection rate $j_{ex}=10^{-2}$, that is,
$j_{in}=j_{out}=5\times 10^{-3}$ in the present assumption.

We can  see in the figures that the photon-photon pair-creation process 
creates a more gap current at around the inner boundary. This is because 
(1) the inward propagating gamma-rays mainly produce the pairs and 
(2) the mean free path is shorter at the inner magnetosphere.  Our result 
confirms the results of our previous calculations (Cheng et al. 2000; 
Takata et al. 2006) and recent 3-D calculation (Hirotani 2015). 

The field aligned electric field separates the electrons and positrons 
created inside the gap, which migrate inward and outward directions, 
respectively, for the inclination angle $\alpha<90^{\circ}$. Since most of 
pairs are created near the inner boundary, the positrons 
 will feel  most of full electric potential drop before escaping from the 
gap outer boundary, while the electrons will feel smaller potential drop between
 the inner boundary and the pair-creation position. As a result, the 
radiation power from the positrons is about factor of ten larger than 
that from the electrons. Figure~\ref{spec} shows the spectra of the 
gamma-rays emitted by outward (solid line) and inward (dashed line) migrating  
particles.  This result is also consistent with 
the previous studies (Takata et al. 2006; Hirotani 2015). 

Figure~\ref{current} shows the  distribution of the electric
current carried
 by the positrons (solid line) and electrons (dashed line) along a  magnetic 
field line for stationary outer gap. The electric field in the gap discharges
 the electrons and positrons and increases the electric current. In the figure, 
the current is constant below low inner  boundary, which is located at 
$r\sim 0.05R_{lc}$, since we assume there is no electric field along the 
magnetic field line between the stellar surface and the inner boundary of the
 outer gap.  In the outer magnetosphere around the light cylinder, the
optical depth of the photon-photon pair-creation process
 is so low that the current density is almost constant 
along the magnetic field line. 

One can estimate the pair-creation mean-free path and multiplicity around the light
cylinder. The mean free path of the photon-photon collision for a gamma-ray
may be  estimated from 
\begin{equation}
\ell_{X\gamma}\sim\frac{1}{(1-\cos\theta_{X\gamma})n_X\sigma_{X\gamma}}
\sim 10^{11}{\rm cm}(1-\cos\theta_{X\gamma})^{-1}(k_BT/80{\rm eV})^{-3} (P/0.1{\rm s})^2,
\end{equation}
where $\theta_{X\gamma}$ is the collision angle, $n_X\sim \sigma_{SB}T^3R_s^2/
R_{lc}^2ck_B$ with $T$ being the temperature of the neutron star surface, 
$R_s$ stellar radius, $\sigma_{SB}$ Stephan-Boltzmann constant and $k_B$ 
Boltzmann constant. In addition, we assume the cross-section
as $\sigma_{X\gamma}=\sigma_T/3$ with $\sigma_T$ being Thomson cross-section.
Optical depth is 
\begin{equation}
\tau_{X\gamma}\sim R_{lc}/\ell_{X\gamma}\sim 
5\times 10^{-3}(1-\cos\theta_{X\gamma})(k_BT/80{\rm eV})^3(P/0.1{\rm s})^{-1},
\end{equation}
which is much smaller than unity for the middle age pulsars and 
the millisecond pulsars.  

A charge particle emit the gamma-rays with a rate of 
\begin{equation}
P_c=\frac{8\pi}{9}\frac{e^2\Gamma}{h R_c}
\sim 2\times 10^6{\rm s^{-1}}(\Gamma/10^7)(R_c/R_{lc})(P/0.1{\rm s}).
\end{equation}
A pair multiplicity by a charge particle accelerated inside the gap 
may be  estimated as 
\begin{equation}
{\cal M}\sim P_c\tau_{X\gamma}R_{lc}/c\sim 150(1-\cos\theta_{X\gamma})(k_BT/80{\rm eV})^3(\Gamma/10^7). 
\end{equation}

\begin{figure}
\includegraphics{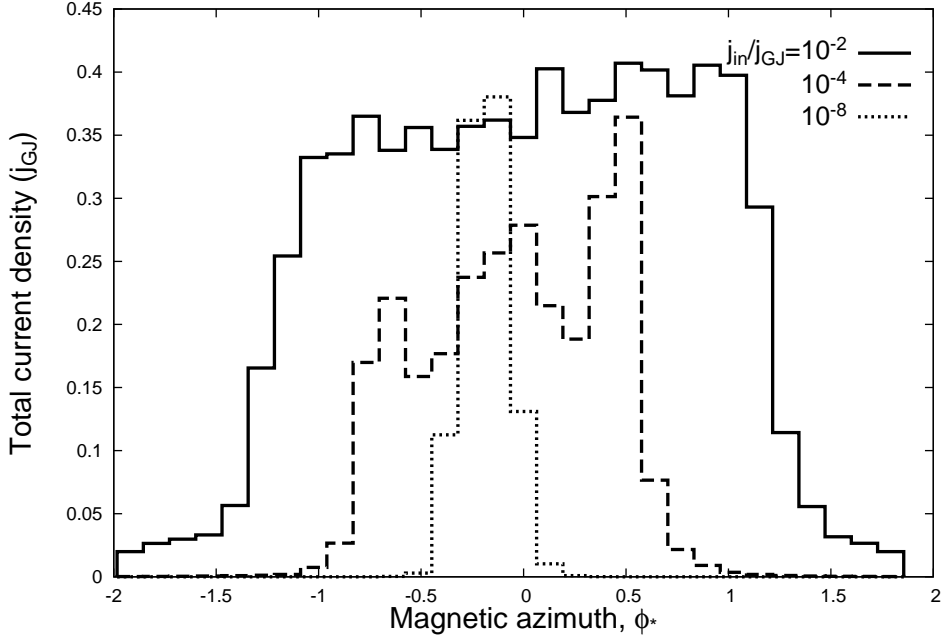}
\caption{Example of  the maximum current density 
 as a function of the  magnetic azimuth. 
The solid, dashed and dotted lines correspond 
to the injection rate $j_{ex}=10^{-2}, 10^{-4}$ and $10^{-8}$, 
respectively.}
\label{cmax}
\end{figure}  

\begin{figure}
\includegraphics{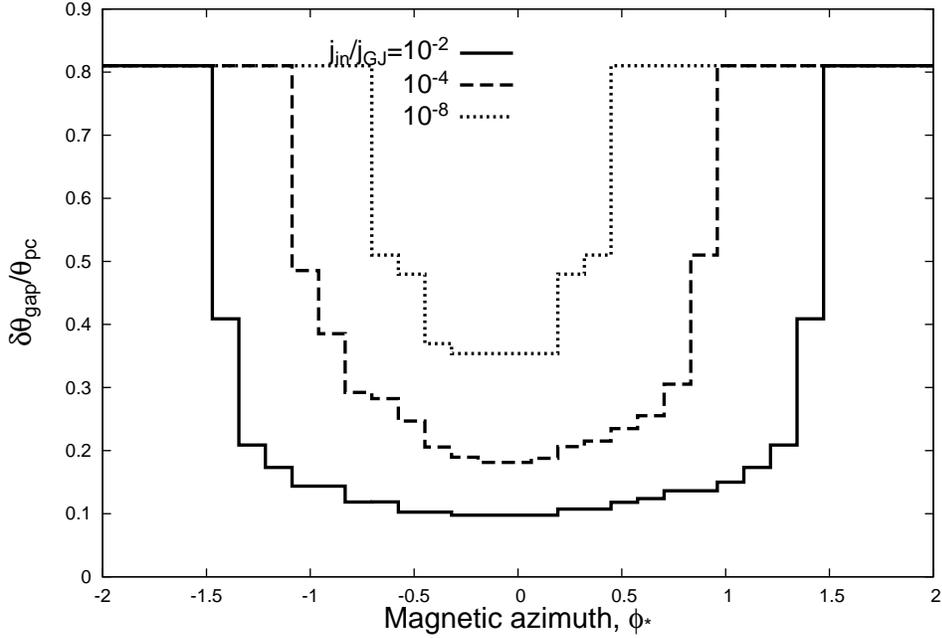}
\caption{Example for the gap thickness (in units of the polar cap size) 
as a function of the
 magnetic azimuth. The solid, dashed and dotted lines correspond 
to the injection rate $j_{ex}=10^{-2}, 10^{-4}$ and $10^{-8}$, 
respectively.}
\label{height}
\end{figure}  
\begin{figure}
\includegraphics{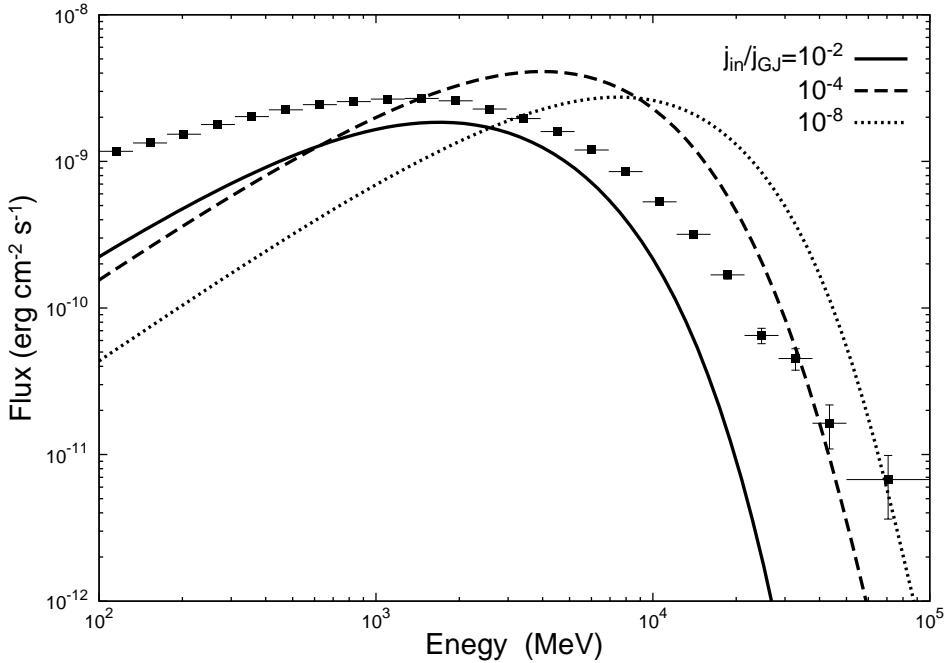}
\caption{Gamma-ray spectra of the Vela pulsar. 
The solid, dashed and dotted lines assume
 the injection rate $j_{ex}=10^{-2}, 10^{-4}$ and $10^{-8}$, respectively. 
The filled boxes show the results of the Fermi
 observations (Leung et al. 2014).}
\label{spec1}
\end{figure}

\begin{figure}
\includegraphics{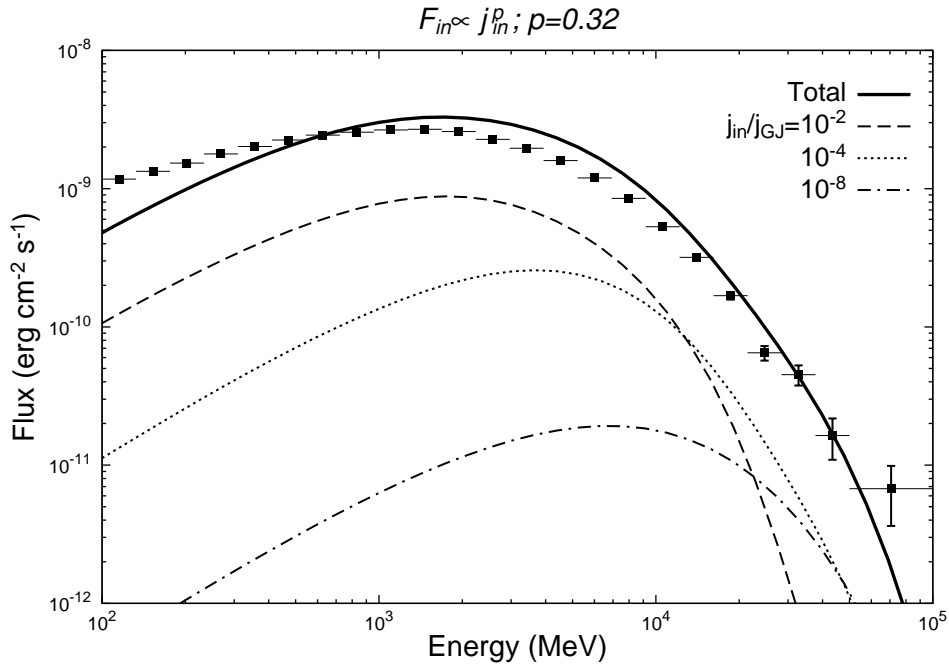}
\caption{The gamma-ray spectrum from the outer gap of the Vela pulsar. 
The solid line shows the gamma-ray spectrum by assuming the power-law 
distribution of the injection rate with a 
index $p=0.32$. The dashed, dotted and dashed-dotted lines show the 
contributions for $j_{ex}=10^{-2}$, $10^{-4}$ and $10^{-8}$,
 respectively. The result is for the inclination angle $\alpha=60^{\circ}$. } 
\label{velaspe}
\end{figure}

\subsection{Gap structure}
Different injection rates produce different 
 outer gap structures and the gamma-ray spectra.  
 Figures~\ref{cmax} and \ref{height} show the distribution of the electric
 current and  the gap thickness, respectively,  as a function of
 the magnetic azimuth.
 In addition, Figure~\ref{spec1} shows the gamma-ray spectrum for 
fixed injection rate. The solid line, dashed line and dotted line in 
the figures show the results for the different injection rate 
$j_{ex}=10^{-2}$, $10^{-4}$ and  $10^{-6}$, respectively. 
We can see in Figure~\ref{cmax} how the azimuthal width of the ``active''
 gap region (large current region) depends on the injection rate; the 
active region is wider for a larger injection rate.  
Figure~\ref{height} and Figure~\ref{spec1} show that  as the injection rate
 increases, the averaged  gap thickness becomes thinner and therefore 
the gamma-ray spectrum becomes softer. 

We can find in Figure~\ref{spec1}, the different injection rates produce 
similar amount of the gamma-ray luminosity. 
 The pulsar electrodynamics implies that the gamma-ray 
luminosity is of order of $L_{\gamma}\sim I\Phi_{nco}$, where 
$I$ is the total current flowing into the gap.
As the injection rate increases,  the total
current becomes larger, while the potential drop, which depends on 
the gap thickness as  $\Phi_{nco}\propto \delta\theta_{gap}^2$, becomes smaller. 
Since these two effects compensate each other, the gamma-ray luminosity 
is insensitive to the injection rate. 

Figure~\ref{spec1} also  shows that  
gamma-ray spectrum for a fixed injection 
rate does not  fit  the observed  spectrum  in 0.1-100GeV
 of the Vela pulsar. With a small injection 
rate $j_{ex}<10^{-4}$, the calculated 
spectrum explains the observed flux level above  ~10GeV, but 
 the predicted spectral slope below 10GeV is steeper than the observed one.
 For the large injection $j_{ex}=10^{-2}$, on the contrary, 
the predicted flux above the cut-off energy decays faster than the observed 
one, and it is difficult to reconcile with  the observed flux above 10GeV.  
With the present  framework of the 3-D calculation, we would expect that 
the superposition of the emissions from the different outer gap 
regions is not the main 
reason for the sub-exponential cut-off behaviour of the Vela pulsar. 

\subsection{Gamma-ray spectrum}
We assume that the observed gamma-ray spectrum is a superposition of
the emissions from various stationary gap structures with various injection 
currents ($j_{ex}$) at the boundaries,
for which we assume power-law distribution of the
injection current (\ref{cdis}), $dN/d{\rm log_{10}}j_{ex} \propto j_{ex}^{-p}$.
We  integrated the emissions of entire outer gap
 regions and used  minimum $\chi$-square method to find the  best fitting
 index, $p$, and normalization for the observed phase averaged spectrum.

 Figure~\ref{velaspe}  compares the best fitting  model spectrum with 
the phase-averaged spectrum  for the Vela pulsar; the solid line
 shows the calculated spectrum  with using the best-fitting 
index $p=0.32$, implying that a larger injection rate dominates 
 in the distribution.  The dashed, dotted and dashed-dotted 
curves in Figure~\ref{velaspe} show 
the contributions for the injection $j_{ex}=10^{-2}$, 
$10^{-4}$ and $10^{-8}$, respectively. As we can see in the figure, 
our model suggests that the emissions from 
the outer gap with smaller injection rates $j_{ex}<10^{-4}$, produces 
observed spectrum above 10GeV, although it's integrated flux in 0.1-100GeV
 energy bands is much smaller than the observed one. We also 
see that the emissions from the outer gap with a larger injection rate
 mainly contributes to  the observed integrated flux, 
but it's spectrum (e.g. dashed line) above 10GeV decays faster than 
the observed spectrum . The cut-off feature of the model 
spectrum (solid line)  is in good agreement  with  the sub-exponential decay of 
the observations. 

Around 100MeV, the model spectrum is steeper than the observed spectrum.
 This may imply that the distribution of the particle injection rates deviates 
from the simple power-law function, 
which has been assumed in the present calculation. 
 Moreover, we have ignored the contributions of inward
 emissions, because the luminosity of the outward
 emissions is about one order of magnitude larger than that of the
 inward emissions, as Figure~\ref{spec} shows. 
 Around 100MeV, however, 
the flux level of the inward emissions is only several factor lower  than
 that of the outward emissions and could contribute to the observed emissions. 
The inward emissions probably contribute to the non-thermal 
X-ray emissions from the Vela pulsar, which shows the multiple (three or four) 
peaks in the X-ray light curve  (Harding et al. 2002; Takata et al. 2008). 

Wang et al. (2010) proposed a two layer outer gap model, which divides 
the outer gap  into two regions, namely, main acceleration region and thin
 screening region around the upper boundary. In the main acceleration 
region,  the electric current density is smaller than Goldreich-Julian 
value and the curvature emissions produce 
GeV gamma-rays. In the screening region, a super Goldreich-Julian current 
screens the field aligned electric field, and the curvature radiation 
produces 100-500MeV gamma-rays. Within the framework of 
 the present calculation method, it is difficult 
to reproduce the stationary outer gap, in which the field  aligned 
electric field is positive-definite, with a super Goldreich-Julian current 
density. A more detail investigation will be necessary  
to explain the observed  emissions around 100MeV of the Vela pulsar. 

\subsection{Light curves}
\begin{figure}
\includegraphics{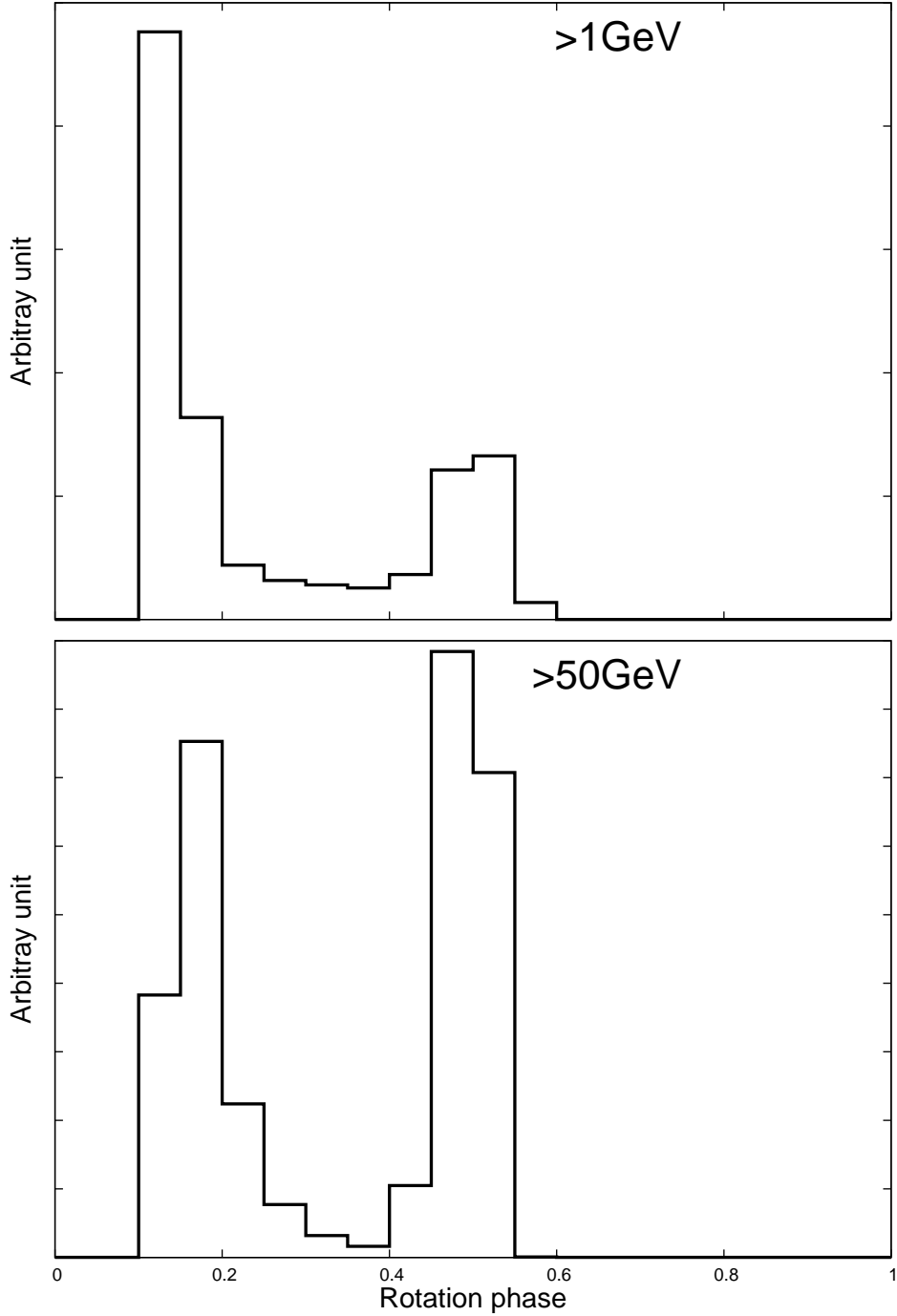}
\caption{Calculated  pulse profiles for $>$1GeV (top) and $>50$GeV (bottom), 
respectively, of the Vela pulsar. The results are for 
the inclination angle $\alpha=60^{\circ}$ and the viewing angle 
$\beta=100^{\circ}$. The south  magnetic  pole ($\phi_*=180^{\circ}$) 
and north magnetic pole ($\phi_*=0^{\circ}$) point towards the observer 
at the rotation phases 0  and 0.5, respectively.} 
\label{light}
\end{figure}
The top and bottom panels in Figure~\ref{light} present 
 the calculated light curves for $>$1GeV and $>50$GeV, respectively,  
of the Vela pulsar. The inclination angle and the viewing angle are 
$\alpha=60^{\circ}$ and $\beta=100^{\circ}$, respectively. In the 
figure, the rotation phase 0 and 0.5 correspond to the times when 
the south  magnetic  pole ($\phi_*=180^{\circ}$) 
and north magnetic pole ($\phi_*=0^{\circ}$), respectively, 
 point towards the observer. 

We find in Figure~\ref{light} that the second peak is more prominent in higher 
energy bands, which is consistent with the observations (Leung et al. 2014). 
In  present  model, the outer gap emissions with a smaller injection rate 
explain the observed  emissions above 10GeV of the Vela pulsar, 
as Figure~\ref{velaspe} shows.  For a smaller injection, the pair-creation 
process in the outer gap produces the pairs  only on the magnetic 
fields around $\phi_*\sim 0$,  as Figure~\ref{cmax} shows. As a result, 
the gamma-rays from the outer gap with a smaller injection rate are 
observed at around orbital phase $\sim 0.5$ in the light curve,
 where is the position of  the second peak.  

We would like to note  that this tendency of energy dependent 
 light curve predicted   by the present model  
is general  behaviour for all gamma-ray pulsars, since 
higher energy photons originate from the outer gap with a smaller injection 
current.  Our model could provide a reason why the second peaks 
of the Fermi-LAT gamma-ray pulsars (Abdo et al. 2012) 
are more prominent in higher energy bands.

\section{Discussion}
\label{disc}
\begin{figure}
\includegraphics{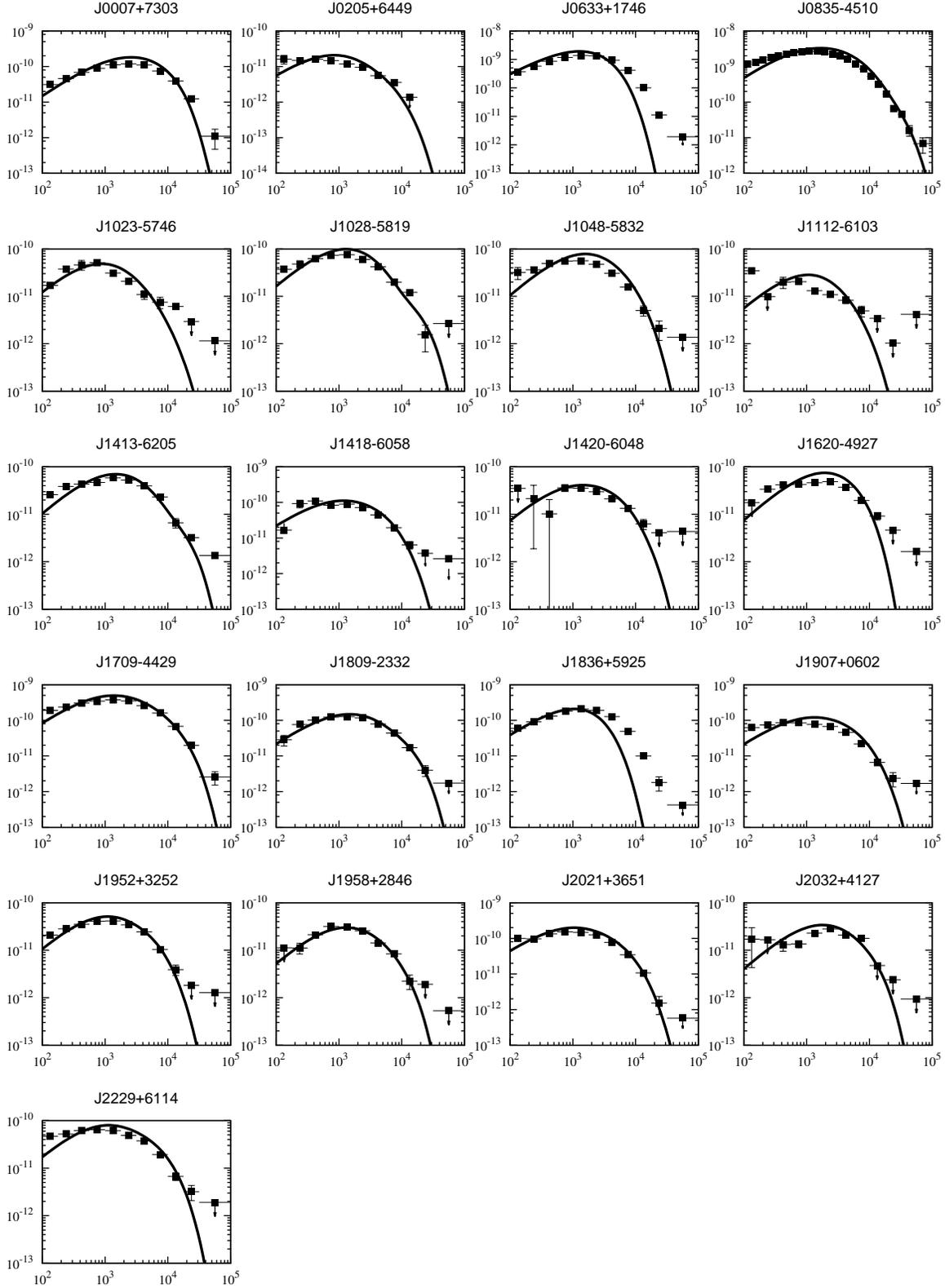}
\caption{The gamma-ray spectra of the young pulsars listed in the Fermi-LAT 
source $>10$GeV. The solid lines show the best fitting model, for which 
the power index $p$ is listed in Table~1.
 The observed spectra were deduced from about 6 years  Fermi observations.
The horizontal and vertical axes represent the 
 energy in unit of MeV and the flux in unit of ${\rm erg~cm^{-2}~s^{-1}}$, 
respectively. }
\label{fit}
\end{figure}

\begin{figure}
\includegraphics{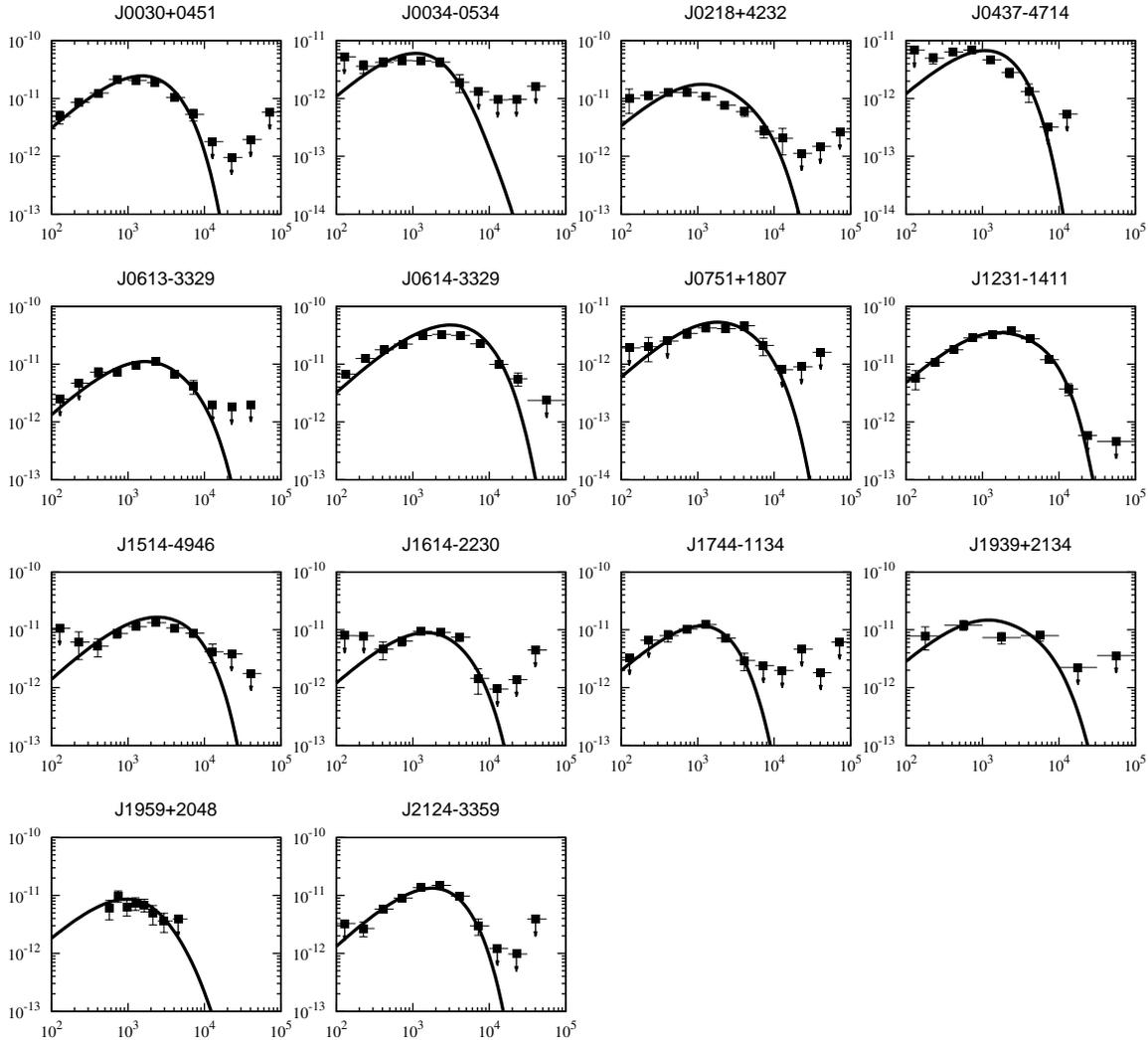}
\caption{The gamma-ray spectra of the millisecond pulsars listed 
the first Fermi-LAT pulsar catalog or the Fermi-LAT $>10$GeV source catalog, 
 expect for original millisecond pulsar J1939+2134 and black widow 
J1959+2048 (Guillemot et al. 2012). The horizontal and vertical 
axes represent the 
 energy in unit of MeV and the flux in unit of ${\rm erg~cm^{-2}~s^{-1}}$, 
respectively. }
\label{fit2}
\end{figure}

The model fitting for various Fermi-LAT pulsars may enable us 
 to discuss how the power-law index $p$ relates  to the spin down parameters. 
Since it is time consuming task 
to investigate for all Fermi-LAT pulsars ($>150$), we 
applied the model to 43 young/middle-age   pulsars (YPSRs) 
and 14 millisecond pulsars (MPSRs). We chose those pulsars from the 
first Fermi-LAT $>$10GeV source catalog (Ackermann et al. 2013) and  
from the first Fermi-LAT pulsar catalog (Abdo et al. 2010b); we exclude the 
Crab pulsar, since its emission process is more complicated.   
We also apply the model to original millisecond pulsar J1939+2138 and 
black widow J1959+2048 (Guillemot et al. 2012) though they are not 
listed in two catalogs. Tables 1-3 summarize the spin down parameters 
for pulsars fitted in this study, and Figures~\ref{fit}-\ref{fit2} compare
 the best fitting model spectra with the Fermi-LAT spectra, for which 
we re-analyzed about 6 years (from 2008 August to 2014 August) 
Fermi data for the pulsars 
listed in first Fermi-LAT $>$10GeV source catalog, while we referred 
the published Fermi spectra (Guillemot et al. 2012; Abdo et al. 2013) for other 
pulsars. To obtain the best fitting model spectra with minimum $\chi$-square method, we use the data points at the center of the errors
 (namely, the values at the filled boxes in each panel
 of Figures~\ref{fit}-\ref{fit2}).
 We ignored the data at the lowest energy bin for fitting 
process since Fermi data at $\sim 100$MeV may contain
 a larger uncertainty.  The last columns  in Tables~1-3 summarize 
the best fitting power-law index $p$ for the distribution of the injection rate 
 
\subsection{Injection current and spin down parameters}
\begin{figure}
\includegraphics{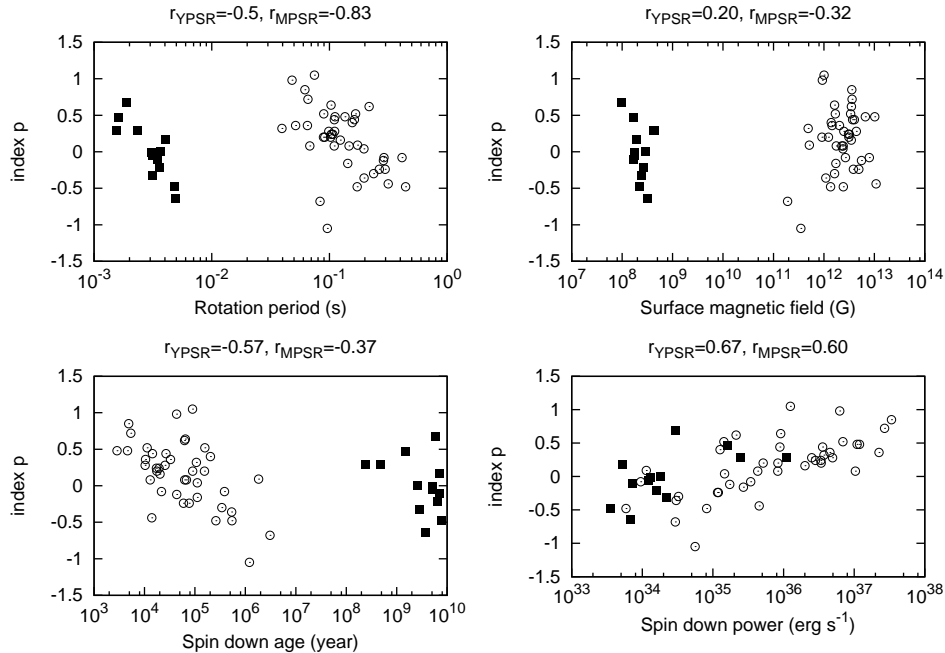}
\caption{The fitting index $p$ vs. the spin down parameters; 
rotation period (top left), surface magnetic field (top right), 
spin down age (bottom left) and spin down power (bottom right). The 
open circles and filled boxes are results for the young pulsars and 
the millisecond pulsars, respectively. In each panel, $r_{YPSR}$ and 
$r_{MPSR}$ show the linear correlation factors for young/middle-age 
 pulsars and millisecond
 pulsars, respectively }
\label{para}
\end{figure}

We investigate how the fitting power-law index $p$ relates to the spin down 
parameters. In Figure~\ref{para}, we plot the fitting power-law 
index $p$ as a function of the spin down parameters, namely, rotation 
period (top-left panel), surface dipole magnetic field (top-right), spin 
down age (bottom-left) and spin down power (bottom-right). In each panel,  
$r_{YPSR}$ and $r_{MPSR}$ are factors of the linear correlation 
for the young/middle-age pulsars  (open circles) 
and millisecond pulsar (filled boxes), respectively.  We find  no correlation 
between the fitting power-law index $p$ and the surface magnetic 
field (top right).  For the rotation period, the correlation is strong 
for the millisecond pulsars but it is relatively weak for the young pulsars. 

With the current samples, 
the correlation between the fitting power-law index and spin
 down power (bottom right in Figure~\ref{para}) 
is relatively stronger for both young pulsars and 
the millisecond pulsars. There is a tendency that  
the fitting index $p$ increases with increasing of the spin down parameter, 
implying a larger current injection is more frequent 
 for the pulsar with a larger
 spin down power.  This tendency 
of the fitting index actually  reflects  the fact that  
the observed spectra below cut-off energy tends to
 be softer (i.e. larger photon index) 
for the pulsar with a larger spin down power (Abdo et al. 2013).  
In other words, our model provides an explanation why the observed 
spectrum below cut-off energy is softer for higher spin down pulsars. 
 
The reason why typical amount of the  injection rate increases with 
 increasing of the spin down power would relate to the increasing 
of the available potential drop with the spin down power. 
The spin down power  relates to the available potential drop on  
the polar cap region $\Phi_{pc}$, namely,
\begin{equation}
L_{sd}\propto B_s^2P^{-4}\propto \Phi_{pc}^{2}. 
\end{equation}
We speculate that the injection currents  
originate from the pair-creation process in 
the acceleration region outside outer gap.  
The pair-creation rate should depend on the 
 available potential drop $\Phi_{pc}$, since the potential drop in the 
acceleration region is proportional to the available potential drop. 
 For  a larger available potential,
 a larger accelerating electric field will arise in the 
acceleration region, and hence more pairs that eventually migrate towards 
the outer gap  will be  created inside and outside acceleration region. 
Hence, we expect that the pulsar with a larger spin down power tends to 
produce a larger particle injection  into the outer gap.   

\subsection{Class II millisecond pulsars}
Pulsars  with gamma-ray peak lagging, aligned with, and preceding the
 radio peak are divided into classes I, II, and III, 
 respectively (Venter et al 2012). For young/middle-age pulsars,
 only Crab and Crab-twin (PSR J0540-6919) in LMC (Fermi-LAT collaboration, 2015)
 show the class II radio/gamma-ray pulse profiles.
 For Fermi-LAT millisecond pulsars, the sources with higher spin down power 
and stronger magnetic field at the light cylinder  in general 
 belong to class II pulsars (Ng et al. 2014). Non-thermal X-ray pulse 
profiles of the class II pulsars show 
similar peak structure and generally align with the gamma-ray 
and  radio peaks. In Table~3, the symbol ``*'' indicates  
the class II millisecond pulsars.

Observed pulsed radio wave  from the class II 
 pulsars  probably originates from the plasma process relating 
to the outer gap accelerator.  We speculate 
that radio emission could be generated above  outer gap region when 
copious  amount of the pairs are created in outer magnetosphere, 
since this could give  a shorter time-scale of  plasma instability 
(Ng et al. 2014).  As we can see in Table~3, our fitting suggests 
that a larger amount of the particles ($p>0$) are injected
 into the outer gap of the class II millisecond pulsars. 
This tendency would be preferable for our speculation for the 
radio emission process of the class II millisecond   pulsars,
 since a larger injection rate can produce a more pairs outside 
the outer gap. We also note that the class II millisecond pulsars 
accompany the giant pulses 
(Romani \& Johnston 2001; Knight et al. 2006), whose phase positions 
are in phase or close to the pulse peaks of normal pulsed radio  emissions. 
Hence the mechanism of the  giant pulses  could relate to the 
 large injection of the particles into the outer gap, 
which results in creation of large amount 
of the pairs. Hence our model  expects 
a correlation between the giant radio pulses and X-ray/gamma-ray emission
 properties.

\subsection{Middle age pulsars; J0633+1746 and J1836+5925}
The middle-age pulsars J0633+1746 (known as Geminga) and J1836+5925 
show that the cut-off behaviours above 10GeV  decay  
slower than pure exponential cut-off, as Figure~\ref{fit} shows.
 Within the present  framework of the calculation 
it is difficult to produce $>$10GeV emissions from the outer gap for 
those two middle-age pulsars, and therefore  there is 
a large discrepancy between the calculated and observed spectra.
  The typical potential drop and the accelerating electric field 
inside the gap are 
of order of $V_{gap}\sim f_{gap}^{2}\times B_sR_s^3/2R_{lc}^2$ and 
$E_{||}\sim V_{gap}/R_c$, respectively, 
where $f_{gap}$ is the ratio of the gap thickness 
to the light cylinder radius at the light cylinder and it is of order of 
unity for middle-age pulsars. The typical Lorentz factor 
of the particles  inside outer  gap for middle-age pulsars is 
$\Gamma\sim (3R_c^2E_{||}/2e)^{1/4}$, which yields the typical energy 
of the curvature photons,
\begin{equation}
E_c=\frac{3}{4\pi}\frac{hc\Gamma^3}{R_{c}}
\sim 0.085{\rm GeV}B^{3/4}_{s,12}P_s^{-7/4}(R_c/R_{lc})^{-1/4},
\end{equation}
where we assumed $f_{gap}=1$. The spin down parameters of 
J0633+1746 and J1836+5925 provides $E_c\sim 1.5$GeV and $\sim 1.2$GeV, 
respectively. This curvature photon energy 
 can explain the position of the spectral cut-off 
around 2GeV, but it is difficult to  reproduce the observed  
emissions above 10GeV for these middle-age  pulsars.  

Vigan$\rm {\grave{o}}$ and Torres (2015) fit the observed spectrum
of the J0633+1746 by parameterizing the magnitude 
of the accelerating electric field in the outer magnetosphere. 
They argued that the observed flux peak position around 2GeV 
requires an  accelerating electric field of order of $E_{||}\sim 10^{7.65}$V/m, 
which corresponds to  a potential drop of order of $\Phi\sim E_{||}R_{lc}
\sim 5\times 10^{14}$V, namely of order of the available potential drop of 
the Geminga pulsar. Their  model phase averaged spectrum also  decays faster
 than the observation above 10GeV.

Takata \& Chang (2009) argued if the last-open field lines could be different 
from the conventional  one that is tangent to the light cylinder. Since 
the magnetic field must be modified  by the rotational and the plasma 
effects in the vicinity of the light cylinder, the size of the polar cap 
could be larger than that of the pure dipole magnetic 
field (Romani 1996; Contopoulos et al. 1999; Gruzinov 2005). 
Since the available potential drop is proportional to the square of the polar 
cap radius, the model flux above 10GeV could increase as increasing of the 
polar cap size.

\subsection{Effects of viewing geometry; PSR J0659+1414}
\begin{figure}
\includegraphics{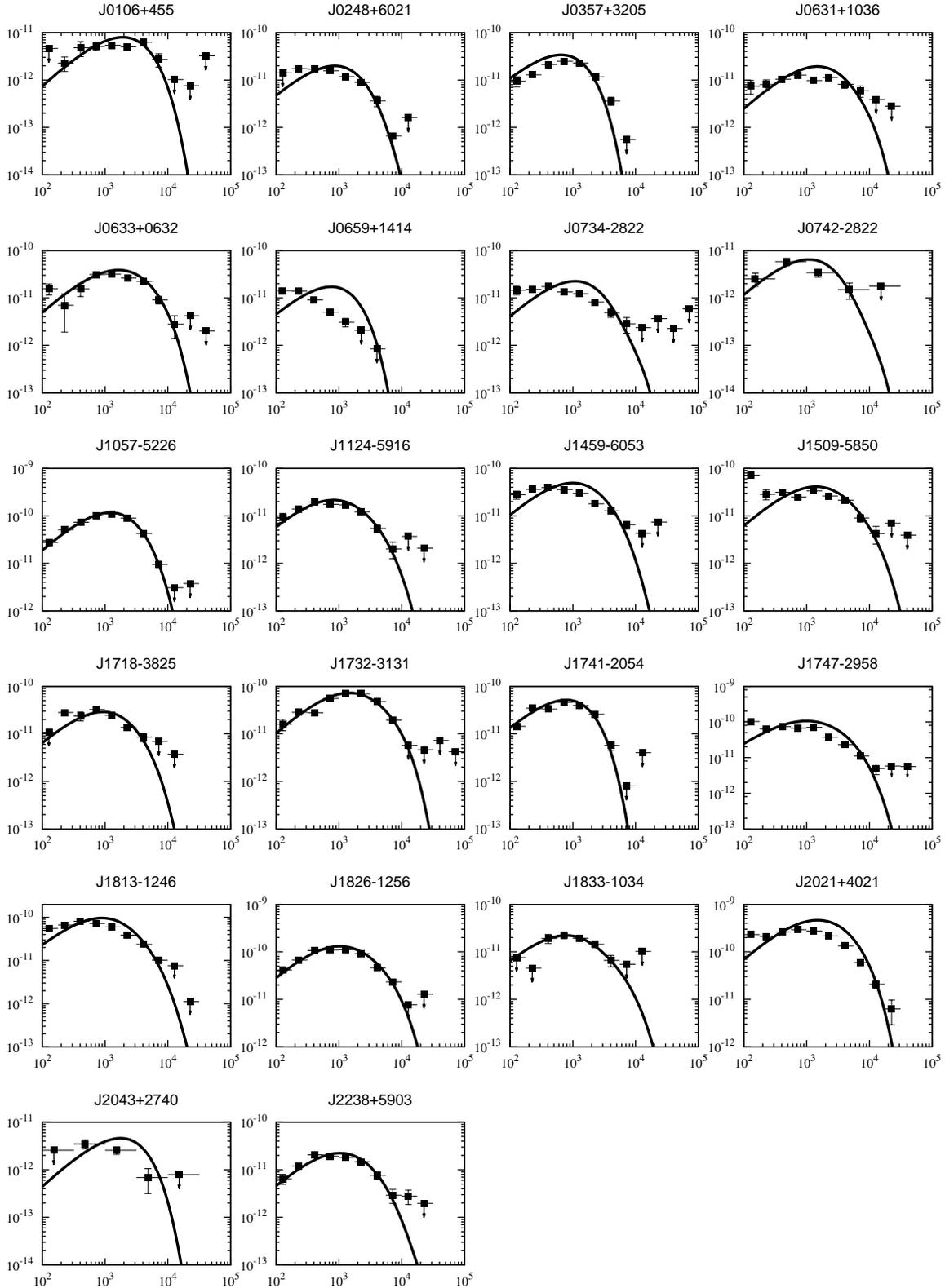}
\caption{The gamma-ray spectra of the young pulsars listed in the first 
Fermi-LAT pulsar catalog (but not listed in the Fermi-LAT $>10$GeV source
 catalog). The observed spectra were taken from Abdo et al. (2013).The horizontal and vertical axes represent the 
 energy in unit of MeV and the flux in unit of ${\rm erg~cm^{-2}~s^{-1}}$, 
respectively.  }
\label{fit1}
\end{figure}

We find in Figure~\ref{fit1} that spectral peak energy ($\sim 0.1$GeV) 
 of PSR J0659+1414 is significantly smaller than the model peak at $\sim$1GeV, 
which corresponds to the  minimum curvature photon energy for 
the pair-creation process,  
$E\sim (m_ec^2)^2/kT_s\sim 2{\rm GeV}(kT_s/0.1{\rm keV})$. For 
PSR J0659+1414, it is likely  that the Earth 
viewing angle cuts through the edge of the gamma-ray beam. 
The outer gap model  predicts that
 more gamma-ray power is released in the direction of $\sim 90^{\circ}$ measured 
from the rotation axis, and hence the Fermi-LAT  has preferentially detected 
pulsars with a larger inclination angle and larger viewing angles
 (Watters \& Romani 2011; Takata et al. 2011). The observation bias would 
 explain the double peak structure in the 
light curves for most of  the Fermi-LAT pulsars.  For PSR~J0659+1414,  
the gamma-ray light curve shows
single peak and furthermore the gamma-ray luminosity divided by the spin 
down power is $\sim 0.006$ (Abdo et al. 2012), 
which is one or two order of magnitudes smaller 
than those of the pulsars with similar spin down power. 
 Hence, we expect that Earth viewing angle of PSR~J0659+1414
 greatly deviates from  $\beta=90^\circ$. 

It could be also possible that we observe the inward emissions for 
PSR~J0659+1414. 
There are  very soft gamma-ray pulsars, which are dim in 
 the Fermi-LAT bands but  bright  sources in hard/soft gamma-ray bands
  (e.g. PSR B1509-54, Wang et al. 2014; Kuiper \& Hermsen 2015). 
We (Wang et al.  2014) suggested that the GeV-quiet soft gamma-ray pulsars  
are  peculiar cases of the viewing  geometry with the Earth viewing 
angle of  $\beta<20^{\circ}$, for which the outward emissions from 
the outer gap is out of line of sight.  
In the model, the inward emissions from the outer gap produce the observed 
spectrum of the GeV-quiet soft gamma-ray pulsars.  With single pulse light 
curve and low efficiency of the gamma-ray emissions, PSR~J0659+1414 could be 
another candidate for which we observe the inward emissions.

In summary, most of Fermi-LAT pulsars show that the spectral behaviour 
above cut-off energy decays slower than pure-exponential function. 
We discussed this sub-exponential cut-off feature 
with non-stationary outer gap accelerator.
For the outer gap accelerator,  the electrons and/or positrons that enter
the gap from the inner and/or outer boundary
control the gap structure and characteristic of the gamma-ray spectrum. 
We found that if rate of the particle injection at the gap boundaries
fluctuate with time,
the gamma-ray spectrum from the outer gap forms a sub-exponential 
cut-off feature. This model predicts that 
the emissions above 10GeV originate from 
a thicker  outer gap with a small injection current,
which also provides a theoretical explanation  why the second peak
is more prominent  in higher energy bands. The observed gamma-ray spectrum 
 below cut-off energy tends to be softer 
for the pulsar with a higher spin down rate. This 
observed tendency is explained if a larger rate of
the particle injection  is more frequent for a higher spin down pulsar. 
The class II millisecond pulsars are very unique
 gamma-ray emitting pulsars. Observed pulsed 
emission in   radio/X-ray/gamma-ray bands and giant radio pulses 
 probably originate from single  site or neighbouring regions 
in outer  magnetosphere. A large injection into the outer gap and 
subsequent pair-creation cascade of the class II millisecond pulsar 
will  create copious pairs outside outer gap, which would enable to 
 develop  a  plasma process for the radio emission.  
We expect that future studies for 
the evolution of the gamma-ray emission properties with the spin down power 
and the correlation of the radio/X-ray/gamma-ray emissions of 
the class II millisecond pulsars advance in understanding for nature 
of the multi-wavelength emission processes in the pulsar magnetosphere.

We express our appreciation to an anonymous referee for useful comments and
suggestions.
We thank C.-Y. Ng, A. H. Kong, C. Y. Hui, P. H. T. Tam, M. Ruderman, and 
S. Shibata for the useful discussions. J.T. and K.S.C. are supported by 
a GRF grant of Hong Kong Government under HKU17300814P. J.T. is supported
by the NSFC grants of China under 11573010. 

\begin{table}
\begin{tabular}{cccccc}
\hline
PSR & $P_s$ & $\tau_{4}$ & $B_{s,12}$ & $L_{sd,35}$ &  p \\
\hline
J0007+7303 & 0.32 & 1.4 & 11 & 4.5 &  -0.44 \\
J0205+6449 & 0.066 & 0.54 & 3.6 & 270 & 0.72 \\
J0633+1746 & 0.24 & 34 & 1.6 & 0.32& -0.3\\ 
J0835-4510 & 0.089 & 1.1 & 3.4 & 69  & 0.52 \\
J1023-5746 & 0.11 & 0.46 & 6.6 & 110  & 0.48 \\
J1028-5819 & 0.091 & 9.0 & 1.2 & 8.3  & 0.2 \\
J1048-5832 & 0.12 & 2.0 & 3.5 & 20  & 0.16 \\
J1112-6103 & 0.065 & 3.3 & 1.5 & 45  & 0.36 \\
J1413-6205 & 0.11 & 6.3 & 1.8 & 8.3  & 0.08 \\
J1418-6058 & 0.11 & 1.0 & 4.4 & 50  & 0.28 \\
J1420-6048 & 0.068 & 1.3 & 2.4 & 100  & 0.08 \\
J1620-4927 & 0.17 & 26 & 1.4 & 0.82  & -0.48 \\
J1709-4429 & 0.10 & 1.8 & 3.1 & 34  & 0.2 \\
J1809-2332 & 0.15 & 6.8 & 2.3 & 4.3  & 0.08 \\
J1836+5925 & 0.17 & 180 & 0.52& 0.11 & 0.09 \\
J1907+0602 & 0.10 & 2.0 & 3.1 & 28 &  0.24 \\
J1952+3252 & 0.039 & 11 & 4.9 & 37 &  0.32 \\
J1958+2846 & 0.29 & 2.2 & 7.9 & 3.4 &  -0.08 \\
J2021+3651 & 0.10 & 1.7 & 3.2 & 34 &  0.24 \\
J2032+4127 & 0.14 & 11 & 1.7 & 2.7 &  -0.16 \\
J2229+6114 & 0.051 & 1.1 & 2.0 & 230 &  0.36 \\
\hline
\end{tabular}
\caption{Young/middle-age  gamma-ray pulsars listed in the Fermi-LAT 
source catalog $>$10GeV (Ackermann et al. 2013) and shown in Figure~\ref{fit}.
 From the left to the right columns, pulsar name (PSR), rotation period ($P_s$) 
in units of second, spin down age ($\tau_4$) in units of $10^4$ years, 
 surface dipole magnetic field ($B_{12}$) in units of $10^{12}$G, spin 
down age $(L_{sd,35}$) in units of $10^{35}{\rm erg~s^{-1}}$ and 
the best fitting power index ($p$) of distribution of the injection current, 
respectively.}
\end{table}

\begin{table}
\begin{tabular}{cccccc}
\hline
PSR & $P_s$ & $\tau_{4}$ & $B_{s,12}$ & $L_{sd,35}$ & p \\
\hline\hline
J0106+455 & 0.083 & 310 & 0.19 & 0.29 &  -0.68 \\
J0248+6021 & 0.22 & 6.2 & 3.5 & 2.1 &  0.62 \\
J0357+3205 & 0.44 & 54 & 2.4 & 20.059 &  -0.48 \\
J0631+1036 & 0.29 & 4.4 & 5.6 & 1.7 &  -0.12 \\
J0633+0632 & 0.30 & 5.9 & 4.9 & 1.2  & -0.24 \\
J0659+1414 & 0.38 & 11 & 4.7 & 0.38  & - \\
J0734-2822 & 0.16 & 20 & 1.4 & 1.3  & 0.4 \\
J0742-2822 & 0.17 & 16 & 1.7 & 1.4  & 0.52 \\
J1057-5226 & 0.20 & 54 & 1.1 & 0.3  & -0.36 \\
J1124-5916 & 0.14 & 0.29 & 10 & 120  & 0.48 \\
J1459-6053 & 0.10 & 6.5 & 1.6 & 9.1  & 0.64 \\
J1509-5850 & 0.089 & 15 & 0.91 & 5.2  & 0.2 \\
J1718-3825 & 0.074 & 9.0 & 1.0 & 13  & 1.05 \\
J1732-3131 & 0.20 & 11 & 2.4 & 1.5  & 0.04 \\
J1741-2054 & 0.41 & 39 & 2.7 & 0.095  & -0.08 \\
J1747-2958 & 0.10 & 2.6 & 2.5 & 25 & 0.28 \\
J1813-1246 & 0.05 & 4.3 & 0.93 & 62  & 0.98 \\
J1826-1256 & 0.11 & 1.4 & 3.7 & 36  & 0.44 \\
J1833-1034 & 0.062 & 0.49 & 3.6 & 340  & 0.85 \\
J2021+4026 & 0.27 & 7.7 & 3.9 & 1.2  & -0.24 \\
J2043+2740 & 0.096 & 120 & 0.35 & 0.56  & -1.05 \\
J2238+5903 & 0.16 & 2.7 & 4.0 & 8.9  & 0.44 \\
\hline
\end{tabular}
\caption{Young/middle-age  gamma-ray pulsars
 listed in the First Fermi-LAT pulsar catalog 
(Abdo et al. 2010b) and but not listed in the Fermi-LAT source catalog 
$>$10GeV. Each column is the same as in Table~1.}
\end{table}
\label{lastpage}
\begin{table}
\begin{tabular}{cccccc} 
\hline
PSR & $P_{-3}$ & $\tau_{9}$ & $B_{s,8}$ & $L_{sd,35}$  & p \\
\hline\hline
J0030+0451 & 4.9 & 7.6& 2.3& 0.035 & -0.48\\
J0034-0534$*$ & 1.9 & 6.0 & 0.98& 0.30 &  0.68\\
J0218+4232$*$ & 2.3 & 0.48 & 4.3 & 2.4 &  0.29 \\
J0437-4715 & 5.8& 1.6& 5.8& 0.12&  {\rm one component}\\
J0613-0200 & 3.1 &  5.1 & 1.7& 0.13 &  -0.32 \\
J0614-3329 & 3.1& 2.8 & 2.4&0.22  & -0.32 \\
J0751+1807 & 3.5 & 7.1& 1.7 & 0.073&  -0.1 \\
J1231-1411 & 3.7& 2.6&2.9&0.18  &0 \\
J1514-4946 & 3.6& 6.4 & 2.6& 0.16&  -0.21\\
J1614-2230 & 3.2& 5.2&1.8& 0.12&  -0.055\\
J1744-1134$*$ & 4.1& 7.2& 1.9& 0.052 & 0.175\\
J1939+2134$*$ & 1.6 & 0.24& 4.1& 12 & 0.29 \\
J1959+2048$*$ & 1.6 &1.5& 1.7&1.6&  0.47\\
J2124-3358 & 4.9& 3.8& 3.2& 0.068&  -0.64\\
\hline
\end{tabular}
\caption{Fermi-LAT millisecond pulsars fitted in this paper. $P_{-3}$ is 
the rotation period in units of millisecond, $\tau_9$ is the spin down age 
in unit of $10^9$ years, and $B_{s,8}$ is the surface dipole magnetic field 
in units of $10^8$G. In the list, the pulsar wind the symbol $*$ represents 
Class II MSP, from which the radio pulse and gamma-ray pulse are in phase.  }
\end{table}

\label{lastpage}


\begin{thebibliography}{}

\bibitem[\protect\citeauthoryear{abdo}{2010}]{ab10a}
Abdo et al., 2010a,  ApJ, 708, 1254
\bibitem[\protect\citeauthoryear{abdo}{2010}]{ab10b}
Abdo et al., 2010b, ApJ, 713, 154
\bibitem[\protect\citeauthoryear{abdo}{2010}]{ab10c}
Abdo et al., 2010c, ApJS, 187, 460
\bibitem[\protect\citeauthoryear{abdo}{2013}]{ab13}
Abdo, A.A. et al., 2012, ApJS, 208,17 
\bibitem[\protect\citeauthoryear{acker}{2013}]{ac13}
Ackermann, M., et al., 2013, ApJS, 209, 34
\bibitem[\protect\citeauthoryear{aharonian}{2012}]{ah12}
Aharonian, F.A., Bogovalov, S.V., Khangulyan, D., 2012, Nature, 482, 507
\bibitem[\protect\citeauthoryear{aleksic}{2011}]{al11}
Aleksi$\rm{\acute{c}}$ et al. 2011, ApJ, 742, 43
\bibitem[\protect\citeauthoryear{aleksic}{2012}]{al12}
Aleksi$\rm{\acute{c}}$ et al. 2012, A\&A, 541, 13
\bibitem[\protect\citeauthoryear{aleksic}{2015}]{al15}
Aleksi$\rm{\acute{c}}$ et al. 2014, A\&AL, 565, 12
\bibitem[\protect\citeauthoryear{aliu}{2008}]{ali08}
Aliu, E., et al., 2008, Sci, 322, 1221
\bibitem[\protect\citeauthoryear{aliu}{2011}]{ali11}
Aliu, E., et al., 2011, Sci, 334, 69
\bibitem[\protect\citeauthoryear{cavero}{2004}]{ca04}
Caraveo, P.A., De Luca, A., Mereghetti, S., Pellizzoni, A., Bignami, G.F.,
 2004, Sci,  305, 376
\bibitem[\protect\citeauthoryear{chen}{2014}]{ch14}
  Chen, A.Y., Beloborodov, A.M., 2014, ApJL, 795, 22
\bibitem[\protect\citeauthoryear{cheng}{2000}]{ch00}
Cheng, K.S., Ruderman, M., Zhang, L., 2000, ApJ, 537, 964
\bibitem[\protect\citeauthoryear{cont}{1999}]{co99}
Contopoulos, I., Kazanas, D., Fendt, C., 1999, ApJ, 511, 351
\bibitem[\protect\citeauthoryear{dau}{1996}]{da96}
Daugherty, J.K., Harding, A.K., 1996, ApJ, 458, 278  
\bibitem[\protect\citeauthoryear{fermi}{2015}]{fe15}
  Fermi-LAT collaborations, 2015, Sci, 13, November, 801-805
\bibitem[\protect\citeauthoryear{crun}{2005}]{cr05}
Gruzinov, A., 2005, PhRv Letter, 94, 1101
\bibitem[\protect\citeauthoryear{gullemot}{2012}]{gu12}
Guillemot, L. et al. 2012, ApJ, 744, 33

\bibitem[\protect\citeauthoryear{harding}{2002}]{ha02}
  Harding, A.K., Strickman, M.S., Gwinn, C., Dodson, R., Moffet, D.,
  McCullock P., 2002, ApJ, 576, 376
\bibitem[\protect\citeauthoryear{harding}{2008}]{ha08}
Harding, A.K., Stern, J.V., Dyks, J., Frackowiak, M., 2008, ApJ, 680, 1378
\bibitem[\protect\citeauthoryear{harding}{2015}]{ha15}
Harding, A.K., Kalapotharakos, C, 2015, ApH, 811, 63
\bibitem[\protect\citeauthoryear{hirotani}{1999}]{hi99}
Hirotani, K., Shibata, S., 1999, MNRAS, 308, 54
\bibitem[\protect\citeauthoryear{hirotani}{2006}]{hi06}
Hirotani, K., 2006, ApJ, 652. 1475
\bibitem[\protect\citeauthoryear{hirotani}{2007}]{hi07}
Hirotani, K., 2007, ApJ, 662, 1173
\bibitem[\protect\citeauthoryear{hirotani}{2015}]{hi15}
  Hirotani, K., 2015, ApJ Letter, 798, 40
\bibitem[\protect\citeauthoryear{Keane}{2013}]{ke13}
 Keane, E. F., 2013, Proceedings of the International Astronomical Union, 291, 295
\bibitem[\protect\citeauthoryear{Knight}{2006}]{kn06} 
Knight, H.S., Bailes, M., Manchester, R.N., Ord, S.M., 2006, ApJ, 653, 580
\bibitem[\protect\citeauthoryear{kramer}{2002}]{kr02}
  Kramer, M., Johnston, S., Van Straten, W. 2002, MNRAS, 334, 523
\bibitem[\protect\citeauthoryear{krause}{1985}]{k85}
Krause-Polstorff, J.,  Michel, F.C., 1985, A\&A 144, 72
\bibitem[\protect\citeauthoryear{kuip}{2015}]{ku15}
Kuiper, L., Hermsen, W., 2015, MNRAS, 449. 3827
\bibitem[\protect\citeauthoryear{leung}{2014}]{le14}
Leung, Gene C. K., Takata, J., Ng, C.W., Kong, A.K.H., Tam, P.H.T., 
Hui, C.Y., Cheng, K.S., 2014, ApJ Letter, 797, 13
\bibitem[\protect\citeauthoryear{lyne}{2010}]{ly10}
Lyne, A., Hobbs, G., Kramer, M., Stairs, I.,  Stappers, B., 2010, Sci, 329, 408	
\bibitem[\protect\citeauthoryear{lyu}{2012}]{lyu12}
Lyutikov, M., Otte, N., McCann, A., MNRAS Letter, 422, 311
\bibitem[\protect\citeauthoryear{ng}{2012}]{ng14}
Ng, C.-Y., Takata, J., Leung, G.C.K.m Cheng, K.S., Philippopoulos, P., 2014, 
ApJ, 787, 167
\bibitem[\protect\citeauthoryear{ozel}{2013}]{oz13}
${\rm\ddot{O}}$zel, F, 2013, RPPh, 76, 6901
\bibitem[\protect\citeauthoryear{romani}{1996}]{ro96}
Romani, R.W., 1996, ApJ, 470, 469
\bibitem[\protect\citeauthoryear{romani}{2001}]{ro01}
Romani, R.W., Johnston, S., 2001, ApJ Letter, 557, 93
\bibitem[\protect\citeauthoryear{shibata}{1995}]{sh95}
 Shibata, S., 1995, MNRAS, 276,  537
\bibitem[\protect\citeauthoryear{shibata}{1991}]{sh91}
 Shibata, S., 1991, ApJ, 378, 239
\bibitem[\protect\citeauthoryear{smith}{2001}]{sm01}
 Smith, I.A., Michel, F.C., Thacker, P.D., 2001, MNRAS, 322, 209
\bibitem[\protect\citeauthoryear{spitkovsky}{2006}]{sp06}
 Spitkovsky, A., 2006, ApJL, 648, 51
\bibitem[\protect\citeauthoryear{taka}{2004}]{ta04}
Takata, J., Shibata, S., Hirotani, K., 2004, MNRAS, 354, 1120
\bibitem[\protect\citeauthoryear{taka}{2006}]{ta06}
Takata, J., Shibata, S., Hirotani, K., Chang, H.-K., 2006, MNRAS, 366, 1310 
\bibitem[\protect\citeauthoryear{taka}{2007}]{ta07}
Takata, J., Chang, H.-K., 2007, ApJ, 670, 677
\bibitem[\protect\citeauthoryear{taka}{2008}]{ta08}
Takata, J., Chang, H.-K., Shibata, S., 2008, MNRAS, 386, 748
\bibitem[\protect\citeauthoryear{taka}{2009}]{ta09}
Takata, J., Chang, H.-K., 2009, MNRAS, 392, 400
\bibitem[\protect\citeauthoryear{taka}{2010}]{ta10}
Takata, J., Wang, Y., Cheng, K.S. 2010, ApJ, 715, 1318
\bibitem[\protect\citeauthoryear{taka}{2011}]{ta11}
Takata, J., Wang, Y., Cheng, K.S., 2011, MNRAS, 414, 2173
\bibitem[\protect\citeauthoryear{taka}{2012}]{ta12}
Takata, J., Cheng, K. S., Taam, Ronald E., 2012, ApJ, 745, 100
\bibitem[\protect\citeauthoryear{vebter}{2012}]{ve12}
  Venter, C., Jphnson, T.J., Harding, A.K., 2012, ApJ, 744, 33
\bibitem[\protect\citeauthoryear{vigano}{2015}]{vi15}
Vigan$\rm {\grave{o}}$, D., Torres, D.F., 2015, MNRAS, 449, 3755
\bibitem[\protect\citeauthoryear{wada}{2011}]{wa11}
Wada, T, Shibata,S, 2011, MNRAS, 418, 612
\bibitem[\protect\citeauthoryear{wada}{2005}]{wa07}
Wada, T, Shibata,S, 2007, MNRAS, 376, 1460
\bibitem[\protect\citeauthoryear{wang}{2014}]{wa14}
Wang, Y., Ng, C.W., Takata, J., Leung, Gene C. K., Cheng, K. S., 2014, MNRAS, 445, 604
\bibitem[\protect\citeauthoryear{wang}{2010}]{wa10}
  Wang, Y., Takata, J., Cheng, K.S., 2010, ApJ, 720, 178
\bibitem[\protect\citeauthoryear{watters}{2011}]{wa11}
  Watters, K., Romani, R.W., 2011, ApJ, 727, 123
\bibitem[\protect\citeauthoryear{yakovlev}{2004}]{ya04}
  Yakovlev, D. G., Pethick, C. J., 2004, ARA\&A, 42, 169
\bibitem[\protect\citeauthoryear{yuki}{2012}]{yu12}
Yuki, S., Shibata, S., 2012, PASJ, 64, 43
\bibitem[\protect\citeauthoryear{zavilin}{2007}]{za07}
Zavlin, V.E., 2007, Ap\&SS, 308, 297
\end{thebibliography}
\end{document}